\documentclass[linenumbers]{aastex631}
\usepackage{threeparttable}

\begin{document}
\nolinenumbers
\shorttitle{The outburst of SWIFT J1727.8--1613 in 2023}
\shortauthors{Peng et al.}
\title{NICER, NuSTAR and Insight-HXMT views to the newly discovered black hole X-ray binary Swift J1727.8--1613}

\author[0000-0002-5554-1088]{Jing-Qiang Peng\textsuperscript{*}}
\email{pengjq@ihep.ac.cn}
\affiliation{Key Laboratory of Particle Astrophysics, Institute of High Energy Physics, Chinese Academy of Sciences, 100049, Beijing, China}
\affiliation{University of Chinese Academy of Sciences, Chinese Academy of Sciences, 100049, Beijing, China}
\author{Shu Zhang\textsuperscript{*}}
\email{szhang@ihep.ac.cn}
\affiliation{Key Laboratory of Particle Astrophysics, Institute of High Energy Physics, Chinese Academy of Sciences, 100049, Beijing, China}

\author[0000-0001-5160-3344]{Qing-Cang Shui\textsuperscript{*}}
\email{shuiqc@ihep.ac.cn}
\affiliation{Key Laboratory of Particle Astrophysics, Institute of High Energy Physics, Chinese Academy of Sciences, 100049, Beijing, China}
\affiliation{University of Chinese Academy of Sciences, Chinese Academy of Sciences, 100049, Beijing, China}
\author[0000-0001-5586-1017]{Shuang-Nan Zhang}
\affiliation{Key Laboratory of Particle Astrophysics, Institute of High Energy Physics, Chinese Academy of Sciences, 100049, Beijing, China}
\affiliation{University of Chinese Academy of Sciences, Chinese Academy of Sciences, 100049, Beijing, China}

\author[0000-0003-3188-9079]{Ling-Da Kong}
\affiliation{Institute f{\"u}r Astronomie und Astrophysik, Kepler Center for Astro and Particle Physics, Eberhard Karls, Universit{\"a}t, Sand 1, D-72076 T{\"u}bingen, Germany}

\author[0000-0001-8768-3294]{Yu-Peng Chen}
\affiliation{Key Laboratory of Particle Astrophysics, Institute of High Energy Physics, Chinese Academy of Sciences, 100049, Beijing, China}

\author[0000-0002-6454-9540]{Peng-Ju Wang}
\affiliation{Institute f{\"u}r Astronomie und Astrophysik, Kepler Center for Astro and Particle Physics, Eberhard Karls, Universit{\"a}t, Sand 1, D-72076 T{\"u}bingen, Germany}
\author[0000-0001-9599-7285]{Long Ji}
\affiliation{School of Physics and Astronomy, Sun Yat-Sen University, Zhuhai, 519082, China}
\author[0000-0002-9796-2585]{Jin-Lu Qu}
\affiliation{Key Laboratory of Particle Astrophysics, Institute of High Energy Physics, Chinese Academy of Sciences, 100049, Beijing, China}
\author[0000-0002-2705-4338]{Lian Tao}
\affiliation{Key Laboratory of Particle Astrophysics, Institute of High Energy Physics, Chinese Academy of Sciences, 100049, Beijing, China}

\author[0000-0002-2749-6638]{Ming-Yu Ge}
\affiliation{Key Laboratory of Particle Astrophysics, Institute of High Energy Physics, Chinese Academy of Sciences, 100049, Beijing, China}
\author[0000-0003-4856-2275]{Zhi Chang}
\affiliation{Key Laboratory of Particle Astrophysics, Institute of High Energy Physics, Chinese Academy of Sciences, 100049, Beijing, China}
\author{Jian Li}
\affiliation{CAS Key Laboratory for Research in Galaxies and Cosmology, Department of Astronomy, University of Science and Technology of China, Hefei 230026, China}
\affiliation{School of Astronomy and Space Science, University of Science and Technology of China, Hefei 230026, China}
\author[0000-0003-2310-8105]{Zhao-sheng Li}
\affiliation{ Key Laboratory of Stars and Interstellar Medium, Xiangtan University, Xiangtan 411105, Hunan, China}

\author{Zhuo-Li Yu}
\affiliation{Key Laboratory of Particle Astrophysics, Institute of High Energy Physics, Chinese Academy of Sciences, 100049, Beijing, China}
\author{Zhe Yan}
\affiliation{University of Chinese Academy of Sciences, Chinese Academy of Sciences, 100049, Beijing, China}
\affiliation{Yunnan Observatories, Chinese Academy of Sciences, Kunming 650216, China}
\affiliation{Key Laboratory for the Structure and Evolution Celestial Objects, Chinese Academy of Sciences, Kunming 650216, China}
\affiliation{Center for Astronomical Mega-Science, Chinese Academy of Sciences, Beijing 100012, China}

%\author{Peng Zhang}
%\affiliation{College of Science, China Three Gorges University, Yichang 443002, China }
%\affiliation{Center for Astronomy and Space Sciences, China Three Gorges University, Yichang 443002, China}

%\author{Yun-Xiang Xiao}
%\affiliation{Key Laboratory of Particle Astrophysics, Institute of High Energy Physics, Chinese Academy of Sciences, 100049, Beijing, China}
%\affiliation{University of Chinese Academy of Sciences, Chinese Academy of Sciences, 100049, Beijing, China}

%\author{Shu-Jie Zhao}
%\affiliation{Key Laboratory of Particle Astrophysics, Institute of High Energy Physics, Chinese Academy of Sciences, 100049, Beijing, China}
%\affiliation{University of Chinese Academy of Sciences, Chinese Academy of Sciences, 100049, Beijing, China}



\begin{abstract}
\nolinenumbers
Swift J1727.8--1613 is a  black hole X-ray binary newly discovered in 2023. We perform spectral analysis with simultaneous Insight-HXMT, NICER and NuSTAR observations when the source was approaching to the hard intermediate state. Such a joint view reveals an additional hard component apart from the normally observed hard component with reflection in the spectrum, to be distinguished from the usual black hole X-ray binary systems. By including this extra component in the spectrum,  we have measured a high spin of $0.98^{+0.02}_{-0.07}$ and an inclination of around $40^{+1.2}_{-0.8}$ degrees, which is consistent with
NICER results reported before.
However, we find that the additional spectral component can not be exclusively determined due to the model degeneracy. Accordingly,  a possible jet/corona configuration is adjusted to account for the spectral fitting with different model trials. The extra component may originate either from a relativistic jet or a jet base/corona underneath a slow jet.

\end{abstract}



\section{Introduction} \label{intro}

A black hole X-ray binary (BHXRB)  consists of a companion star and a black hole (BH). BHXRBs can be classified into two categories: low-mass X-ray binaries, where the companion star has a low mass ($\lesssim 1M_\odot$), and high-mass X-ray binaries, where the companion star has a higher mass. Additionally, BHXRBs can be further classified as either persistent or transient sources, depending on their behavior \citep{2016Tetarenko, 2018Sreehari}.
For transient sources that remain in a quiescent state with a very low accretion rate for a long time, the accretion disk is a standard Shakura-Sunyaev disk, and the inner radius of the disk is truncated at a distance far from the BH. At this time, the accretion rate of the disk is very low, and the temperature is also very low \citep{1973Shakura}.

Due to the low accretion rate, the temperature of the disk remains relatively low, and the material within the disk predominantly exists in a neutral state. As the accreted material accumulates in the disk, its temperature gradually increases to the ionization temperature of hydrogen ($ \sim 10^5$ K), causing ionization of hydrogen. This ionization process triggers thermal and viscous instabilities, which lead to an outward transfer of angular momentum, inward motion of the material, and an increase in the rate of accretion, producing an X-ray outburst \citep{1995Cannizzo, 2001Lasota,2011Belloni,2016Corral-Santana}.

BH X-ray outbursts are commonly represented by the hardness-intensity diagram (HID), which allows for their classification into different spectral states. These states include the  Low/Hard States (LHS), High/Soft States (HSS), and
Intermediate States (IMS)  based on the position of the trajectory of the outbursts on the HID \citep{2005Belloni,2009Motta}.
The characterization varies from one spectral state to another. In the LHS, the emission is mainly from the corona/jet, the non-thermal component dominates, and the photon index $\Gamma$ is around 1.5. Hard tails are sometimes present in the spectrum usually thought to be related to the jet \citep{2016Reig,2021Kong}. In the soft state, non-thermal radiation from the disk dominates, and the energy spectrum is softer, with a higher spectral index around 2.1--3.7 \citep{1997Esin,2006McClintock}.
The IMS serves as an intermediary between the LHS and HSS and can be further divided into hard and soft intermediate states \citep{2005Belloni, 2005Homan}.

For BHs, their mass and spin are two crucial parameters that play a significant role in understanding their origin and evolution. The spin of a BH provides valuable insights into how it interacts with its surrounding accretion environment and influences various processes such as jet production.
There are typically two methods used for measuring the spin of BHs. The first method is relativistic reflection, which involves estimating the spin by analyzing the profile of iron lines and the Compton hump near $\sim$ 20 keV \citep{2006Brenneman,2009Miller}. The second method is through continuum fitting, which requires the inner accretion disk to be in the innermost stable circular orbit (ISCO) \citep{1997Zhang,2005Li}. 

Swift J1727.8--1613 was first detected by the SWIFT/BAT telescope on August 24, 2023. with further observations in X-ray, radio, and optical wavelengths, it was subsequently identified and confirmed as a low-mass BH X-ray binary \citep{2023Nakajima,2023O'Connor,2023Castro-Tirado,2023Miller}. 
In the hard X-ray band, there are prominent Quasi-Periodic Oscillation (QPO) frequencies observed within the range of 0.46--0.88 Hz \citep{2023Palmer}. \cite{2023Draghis} obtained an interstellar absorption of 0.41$\pm0.01$ $\times$ $10^{22}$ $\rm cm^{-2}$ by fitting the spectrum of NICER. Additionally, a relativistically broadened Fe K emission line was detected at the 5--7 keV range;
for the first time the spin of the BH and the inclination of the system were reported as $a=0.995^{+0.001}_{-0.004} $ and $\theta=47.9\pm0.03$ degrees, respectively.
Furthermore, in the residuals of the fit, an absorption feature near 7 keV was observed, suggesting the presence of an accretion disk wind.

In this paper, we analyze the outburst data of Swift J1727.8--1613 collected simultaneously by snapshots of Insight-HXMT, NICER and NuSTAR, and thus present the first views to the spectral properties of Swift J1727.8--1613 in a rather broad energy band.  In Section \ref{obser}, we describe the observations and data reduction. The detailed results are presented in Section \ref{result}. The results are discussed and the conclusions are presented in Section \ref{dis}.

\section{Observations and Data reduction}
\label{obser}

\begin{table}[htbp]
    \centering
		\caption{NICER, Insight-HXMT and NuSTAR  observations of Swift J1727.8--1613  during the 2023 outburst. }
		\begin{tabular}{cccccc}
		%\begin{tabular}{|c|m{2cm}|}
		  \hline
		   \hline
		   NICER & Observed date & Exposure Time \\ObsID& (MJD) & (s) \\ 
       6703010101 &  60185.55 &1528   \\ 
		  \hline \hline
         NuSTAR &  Observed date & FPMA exposure time & FPMB exposure time
         \\ ObsID& (MJD) & (s) & (s) 
         \\ \hline
       90902330002 &  60185.42 &923& 1012 \\

     \hline \hline
         Insight-HXMT &  Observed date & LE exposure time & ME exposure time
         \\ ObsID& (MJD) & (s) & (s) 
       \\ \hline
       P051436300803 & 60025.22& 284.3 & 964.2 \\ 
    \hline
        \label{observ} &     

    \end{tabular}

\end{table}

\subsection{Insight-HXMT}

Insight-HXMT is China's first X-ray astronomy satellite and was successfully launched on 2017 June 15 \citep{2014Zhang, 2018Zhang, 2020Zhang}.
It carries three scientific payloads that enable observations across a wide range of energy bands (1--250 keV).
The three scientific payloads: the low energy X-ray telescope (LE, SCD detector, 1--15 keV, 384 $\rm cm^{2}$,
\citealt{2020Chen}), the medium energy X-ray telescope (ME, Si-PIN detector, 5--35 keV, 952 $\rm cm^{2}$, \citealt{2020Cao}), and the high energy X-ray telescope (HE, phoswich NaI(CsI), 20--250 keV, 5100 $\rm cm^{2}$, \citealt{2020Liu}); note that the area of each telescope above is its total geometrical area.

Insight-HXMT started to observe Swift J1727.8--1613  on August 25, 2023. We take an observation that was carried out almost simultaneously with the NuSTAR observation (see table \ref{observ}) when the source evolved to approach the intermediate state (see Figure \ref{HID}). 
We extract the data from ME and HE using the Insight-HXMT Data Analysis software {\tt{HXMTDAS v2.05}}. LE data are not included due to the temporary calibration issues. Since the LE is SCD detector, there is a pile-up effect which is in the process of being calibrated.
The data are filtered with the  criteria recommended by the Insight-HXMT Data Reduction Guide {\tt v2.05}\footnote[1]{{http://hxmtweb.ihep.ac.cn/SoftDoc/648.jhtml}}
{\tt Xspec v12.13.1}\footnote[2]{{https://heasarc.gsfc.nasa.gov/docs/xanadu/xspec/index.html}} is used to perform analysis of spectrum.
Due to poor calibration, the energy bands considered for spectral analysis are ME 10--28 keV and HE 28--140 keV. One percent systematic error is added to data, and errors are estimated via  Markov Chain Monte-Carlo (MCMC) chains with a length of 20000.

\subsection{NuSTAR}

The Nuclear Spectroscopic Telescope Array (NuSTAR) is the first mission to use focusing telescopes to image the sky in the high-energy X-ray (3 - 79 keV) region of the electromagnetic spectrum, which was launched at 9 am PDT, June 13, 2012 \citep{2013Harrison}. The NuSTAR observation carried out simultaneously with Insight-HXMT (see Table \ref{observ}) is adopted to investigate the spectral properties of Swift J1727.8--1613 in a broad energy band. 
We extract NuSTAR-filtered data using the standard pipeline program {\tt nupipeline}, 
Since Swift J1727.8--1613  is very bright we make the parameter statusexpr="(STATUS==b0000xxx00xxxx000)\&\&(SHIELD==0)" followed by using nuproduct to extract the light curves and spectrum. The spectrum and light curves were extracted from a 120$''$ circle region centered on the source and the background was generated from a 60$''$ circle region away from the source. We adopt both FPMA and FPMB data for spectral analyses.

\subsection{NICER}
The X-ray TIming Instrument (XTI) of the Neutron Star Interior Composition Explorer (NICER) is an International Space Station (ISS) payload, which was launched by the Space X Falcon 9 rocket on 3 June 2017 \citep{2016Gendreau}. NICER has a large effective area and high temporal resolution in soft X-ray band (0.2--12 keV), which may well fit the black body component at low temperatures.

The observation from NICER that was observed simultaneously with NuSTAR and Insight-HXMT (see Table \ref{observ})  extends the investigation of the spectral properties of Swift J1727.8--1613  to lower energy bands.
NICER data are reduced using the standard pipeline tool {\tt \textit{NICER}l2\footnote[3]{https://heasarc.gsfc.nasa.gov/lheasoft/ftools/headas/\textit{nicer}l2.html}}. 
We extract light curves using {\tt \textit{nicer}l3-lc\footnote[4]{https://heasarc.gsfc.nasa.gov/docs/software/lheasoft/ftools/headas/\textit{nicer}l3-lc.html}} in 1--10 keV.
We use {\tt \textit{nicer}l3-spect\footnote[5]{https://heasarc.gsfc.nasa.gov/docs/software/lheasoft/help/\textit{nicer}l3-spect.html}} to extract the spectrum, with "{\tt nibackgen3C50\footnote[6]{https://heasarc.gsfc.nasa.gov/docs/\textit{NICER}/analysis\_{}threads/background/}}" model to estimate the
background for the spectral analysis;
 {\tt \textit{NICER}l3-spect} also applies the systematic error using niphasyserr automatically.
For the fitting of the spectrum, we choose an energy range of 1--10 keV, because NICER below 1 keV has significant residual due to calibration problems.

\section{Results}
\label{result}

\subsection{ Hardness ratio}

\begin{figure}
	\centering
	\includegraphics[angle=0,scale=0.1]{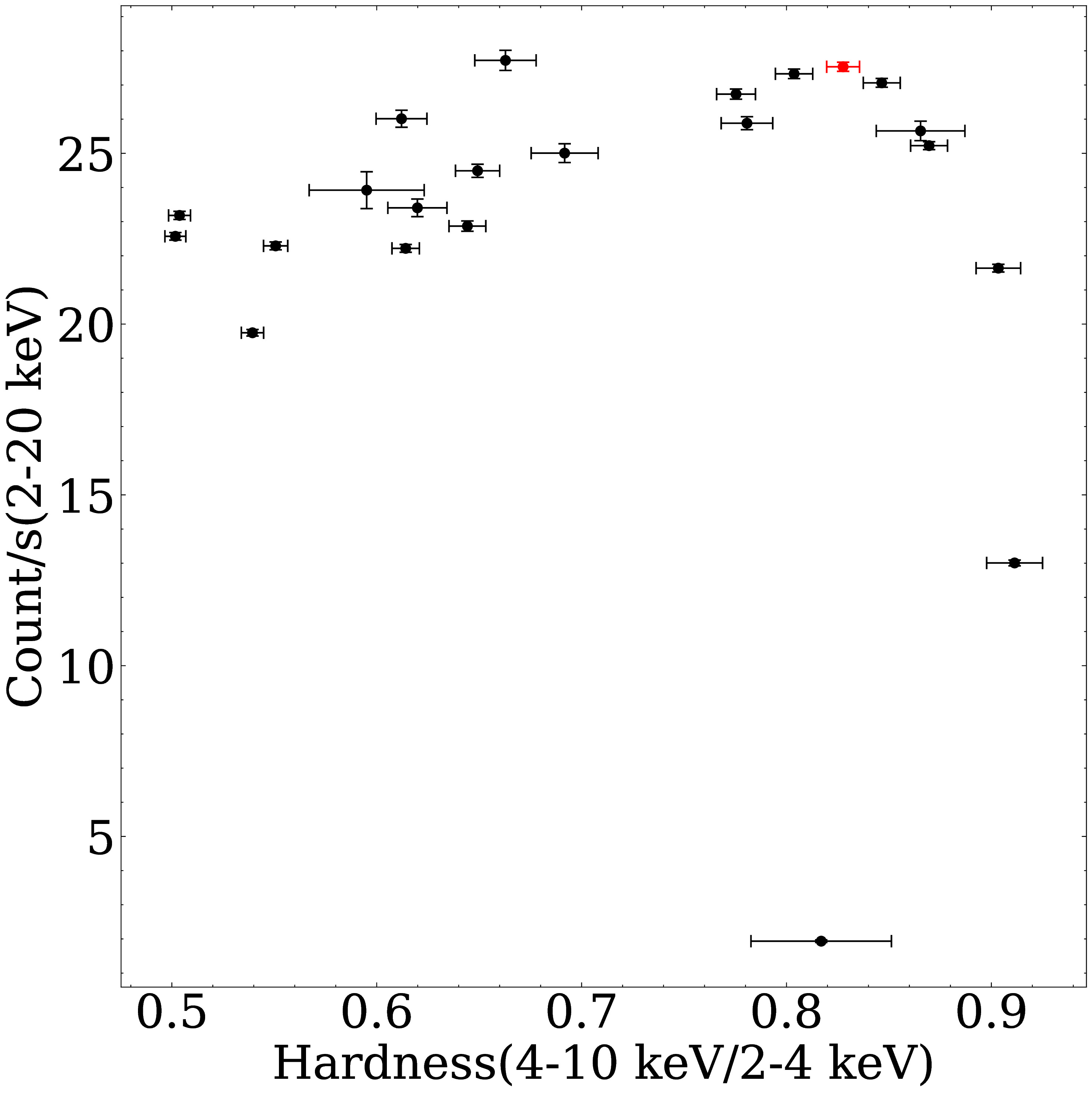}
	\caption{The MAXI hardness-intensity diagram of Swift J1727.8--1613 with time bin 1 d, where the hardness is defined as the ratio of 4--10 keV to 2--4 keV count rate. The red dot is the time of the observation in Table \ref{observ}.}
	\label{HID}
\end{figure}

As shown in Table \ref{observ}, Swift J1727.8--1613 was observed on MJD 60185 by NuSTAR with ObsID 90902330002, by NICER with ObsID 6703010101, and by Insight-HXMT ME and HE with ObsID P061433800302.
To investigate the timing of this observation during the Swift J1727.8--1613 outburst, we construct the MAXI hardness-intensity diagram of Swift J1727.8--1613 with time bin 1 d. As shown in Figure \ref{HID}, the hardness is defined as the ratio of 4--10 keV to 2--4 keV count rate. The red dot is the time of joint NuSTAR, NICER and Insight-HXMT observations. The flux of Swift J1727.8--1613 exhibits a rapid increase, and the observation corresponds to the end of the hard state of the outburst.

\subsection{The spectral analysis}
	\label{spec}

\begin{figure}
	\centering
	\includegraphics[angle=0,scale=0.5]{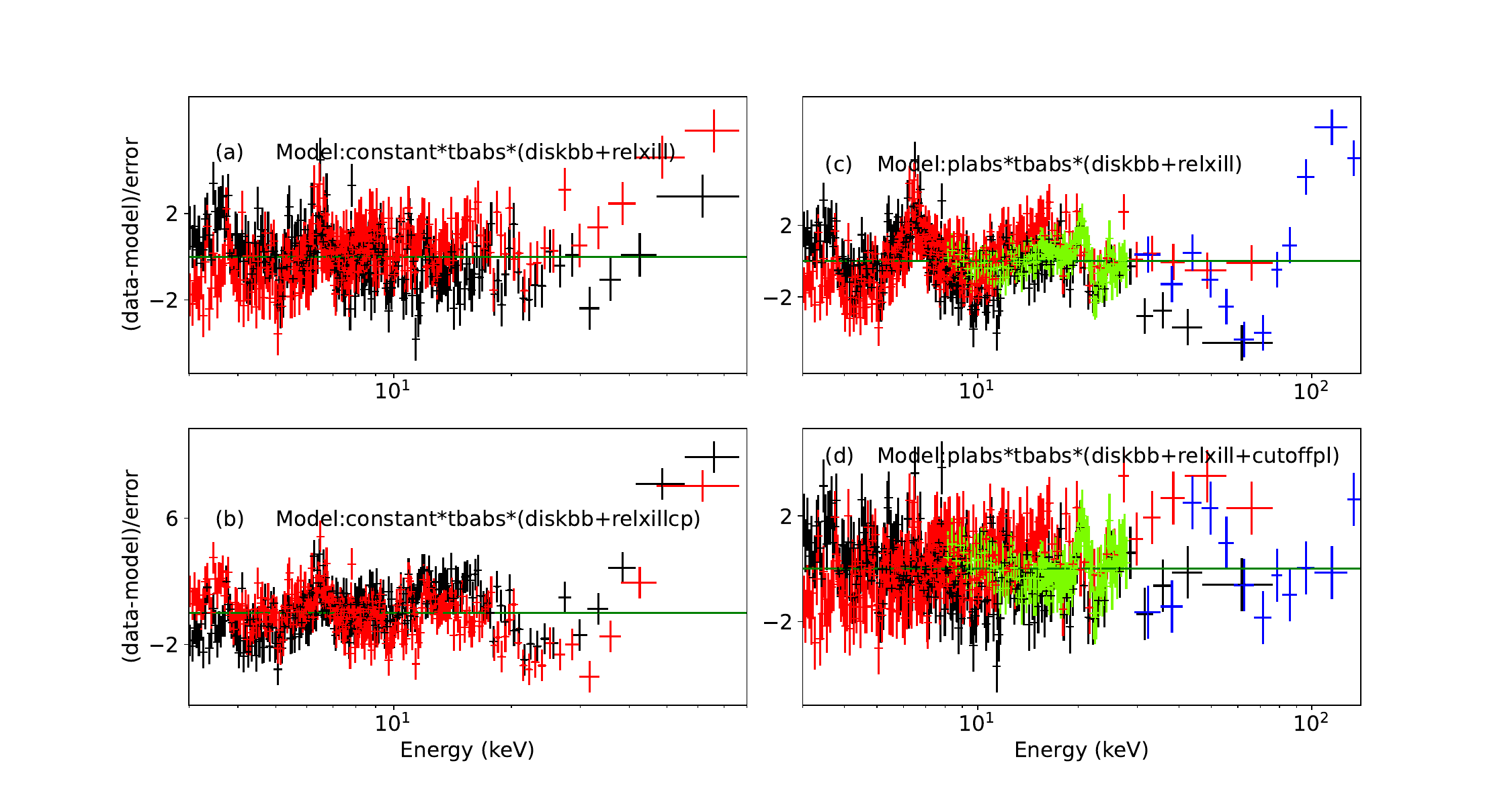}
	\caption{Spectral fitting residuals for different models. The black, red, green and blue points represent spectral residuals from NuSTAR FPMA, FPMB and Insight-HXMT ME, HE respectively. }
	\label{residual}
\end{figure}

\begin{figure*}
\centering

\begin{minipage}[t]{0.45\linewidth}
\centering
\includegraphics[angle=0,scale=0.35]{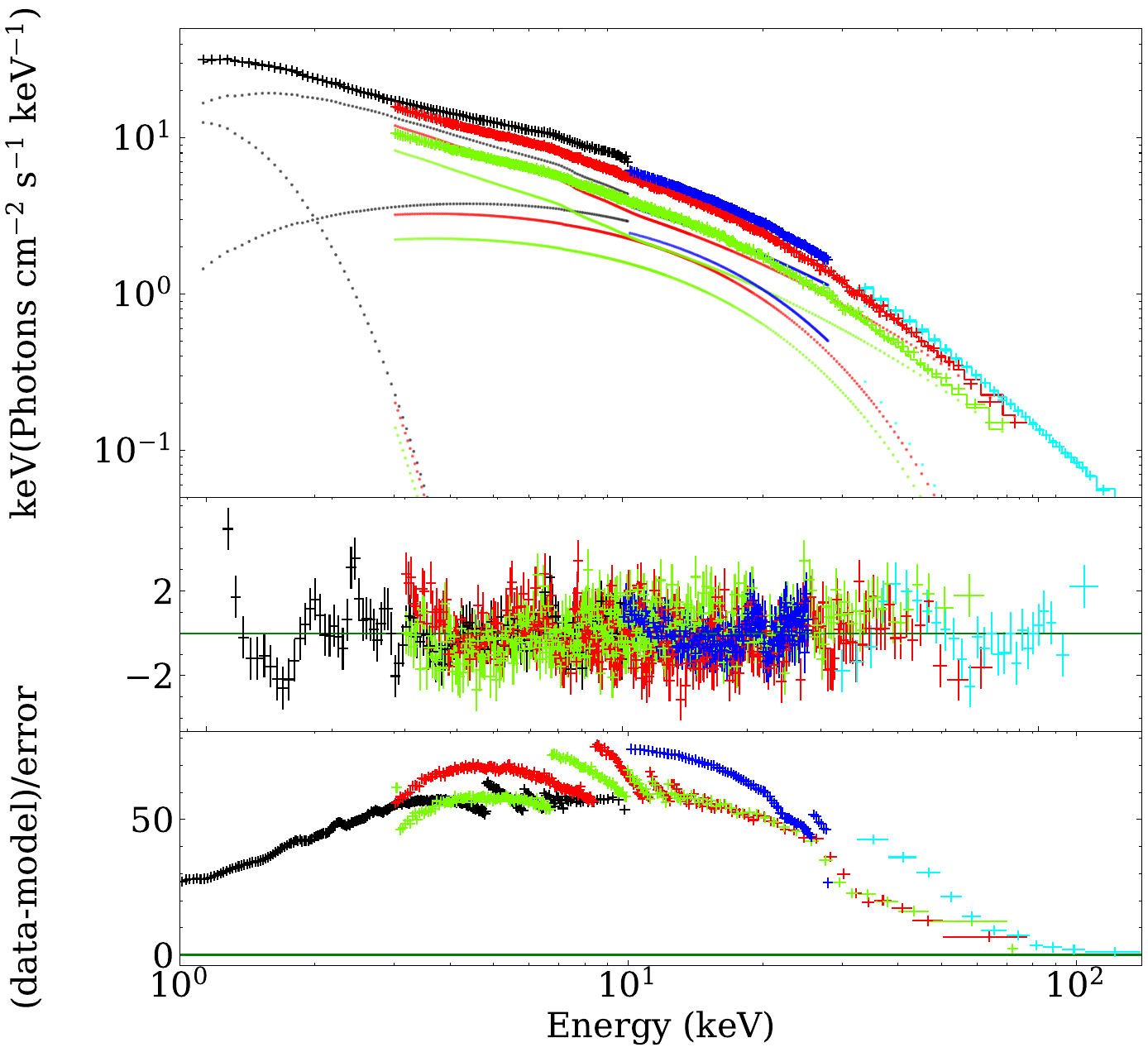}
%\caption{fig1}
\end{minipage}%
\hfill
\begin{minipage}[t]{0.45\linewidth}
\centering
\includegraphics[angle=0,scale=0.35]{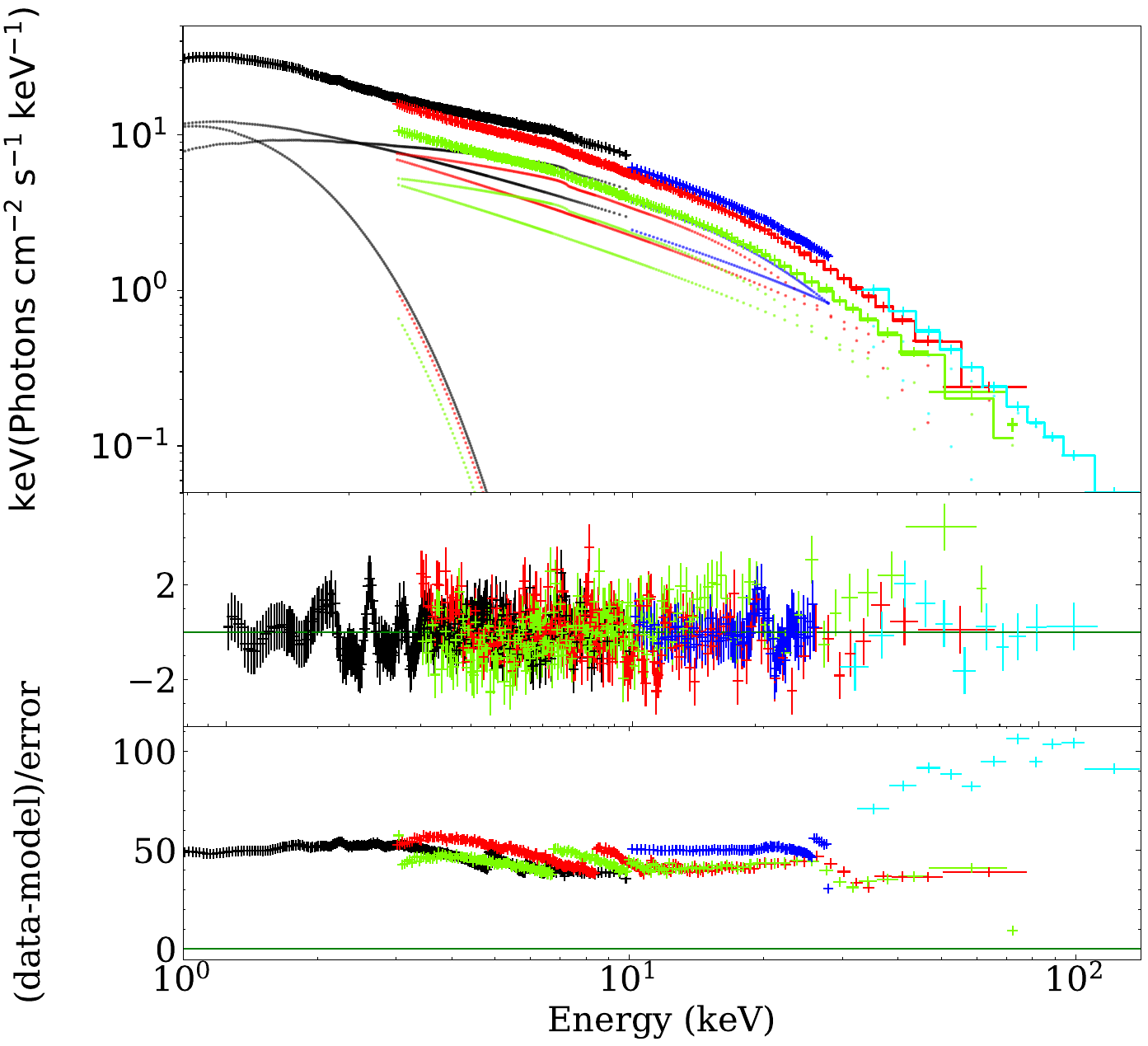}
%\caption{fig2}
\end{minipage}
\centering
\caption{The simultaneous broadband spectrum of Swift J1727.8--1613 is observed from NICER (black) NuSTAR/FPMA (red), NuSTAR/FPMB (green), Insight-HXMT/ME (blue), and Insight-HXMT/HE (cyan).
Left panel: The spectrum of model 1, with a relatively high $\Gamma$ and $E_{\rm cut}$ for the reflection component. The bottom panel is the residual of model 1 after removing cutoffpl.
Right panel: The spectrum of model 1, with a relatively low $\Gamma$ and $E_{\rm cut}$ for the reflection component. The bottom panel is the residual of model 1 after removing cutoffpl.
}
\label{fit}
\end{figure*}

\begin{figure*}
\centering

\begin{minipage}[t]{0.45\linewidth}
\centering
\includegraphics[angle=0,scale=0.35]{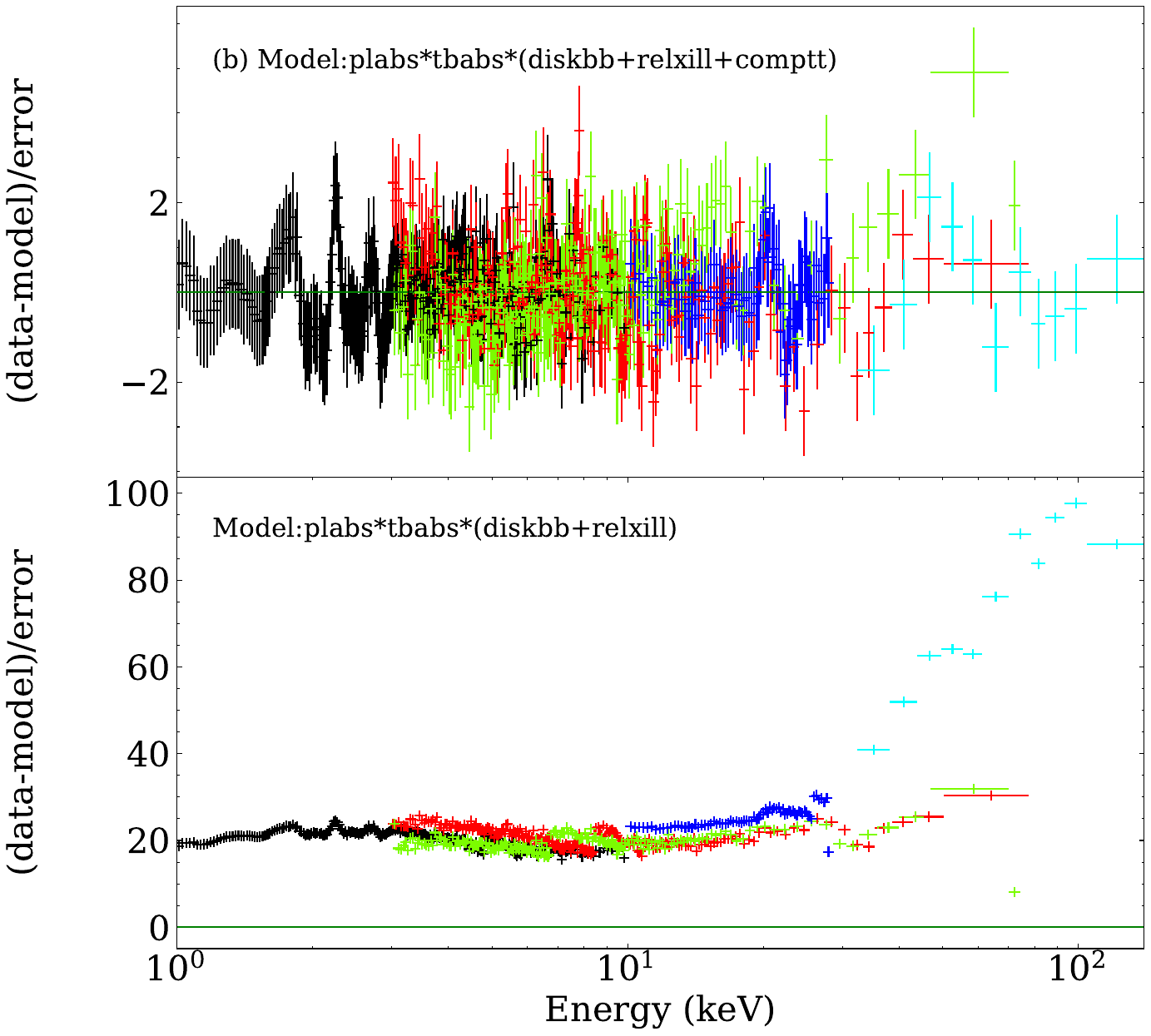}
%\caption{fig1}
\end{minipage}%
\hfill
\begin{minipage}[t]{0.45\linewidth}
\centering
\includegraphics[angle=0,scale=0.35]{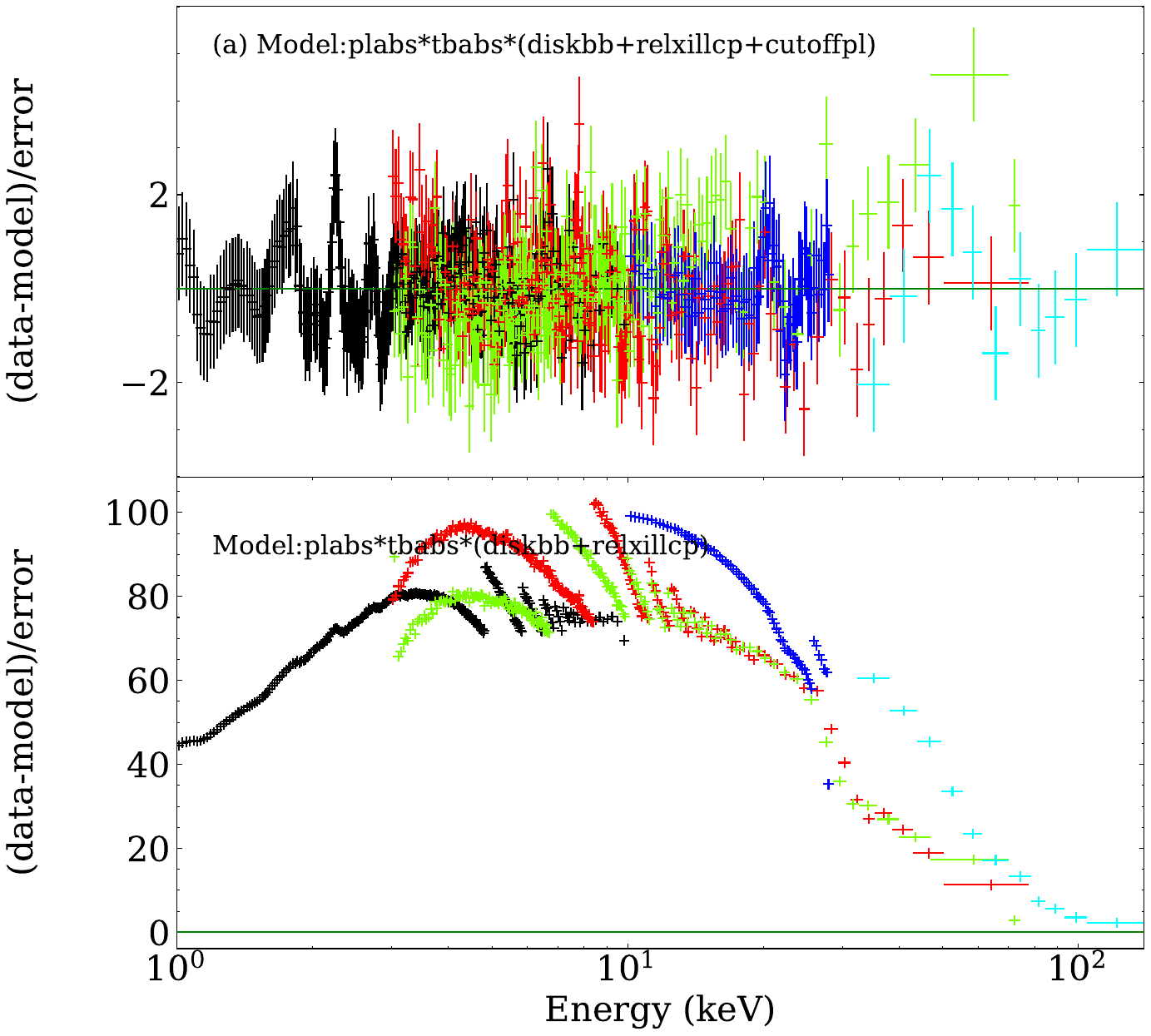}
%\caption{fig2}
\end{minipage}
\centering
\caption{The simultaneous broadband spectrum of Swift J1727.8--1613 is observed from NICER (black) NuSTAR/FPMA (red), NuSTAR/FPMB (green), Insight-HXMT/ME (blue), and Insight-HXMT/HE (cyan).
Left panel: Spectral fitting residuals for model 2. The bottom panel is the residual of model 2 after removing compTT.
Right panel: Spectral fitting residuals for model 3. The bottom panel is the residual of model 3 after removing cutoffpl.
}
\label{re}
\end{figure*}

\begin{table}[htbp]
    \centering
    \renewcommand{\arraystretch}{1.5}
     \setlength{\tabcolsep}{5pt}
		\caption{The results of spectral fitting the NICER+NuSTAR+Insight-HXMT data for Model M1.}
		\resizebox{8cm}{!}{
		\begin{threeparttable}
	\begin{tabular}{cccc}
		%\begin{tabular}{|c|m{2cm}|}
		
		  \hline
		   \hline
        Model & Parameter &M1&$\rm M1^{*}$
       \\ \hline
tbabs& $N_{\rm H}[10^{22} \rm cm^{-2}]$&$0.18^{+0.007}_{-0.004}$&$0.18^{+0.009}_{-0.008}$\\
\hline
diskbb&$T_{\rm in}$&$0.45_{-0.005}^{+0.008}$&$0.45_{-0.009}^{+0.006}$ \\
&norm[$10^{4}$]&$6.33_{-0.60}^{+0.37}$&$6.42_{-0.51}^{+0.75}$ \\
\hline
relxill&$a$&$0.98_{-0.06}^{+0.02}$&$0.98_{-0.07}^{+0.02}$ \\
&$i$ [\textdegree]&$41.0_{-1.14}^{+2.04}$&$40.2_{-0.59}^{+0.90}$\\
&$\Gamma$&$1.72_{-0.02}^{+0.01}$&$1.01_{-0.007}^{+0.026}$ \\
&logxi&$3.16_{-0.11}^{+0.10}$&$3.09_{-0.11}^{+0.07}$\\
&$A_{fe}$&$1.77_{-0.33}^{+0.16}$&$2.50_{-1.24}^{+0.85}$\\
&$E_{\rm cut}$[keV]&$59.8_{-1.93}^{+2.44}$&$12.07_{-1.09}^{+0.20}$\\
&ref\underline{ }frac&$0.36_{-0.03}^{+0.05}$&$0.24_{-0.04}^{+0.06}$\\
&norm&$0.28_{-0.02}^{+0.01}$&$0.23_{-0.006}^{+0.009}$\\
&flux[$10^{-7}$~erg~s$^{-1}$~cm$^{-2}$]&$1.38_{-0.01}^{+0.02}$&$0.95_{-0.02}^{+0.03}$\\
\hline
cutoffpl&PhoIndex&$1.02_{-0.02}^{+0.02}$&$1.73_{-0.01}^{+0.009}$\\
&High$E_{\rm cut}$[keV]&$11.45_{-0.24}^{+0.14}$&$58.27_{-1.39}^{+1.21}$\\
&norm&$10.53_{-0.53}^{+0.45}$&$18.66_{-0.85}^{+0.67}$\\
&flux[$10^{-7}$~erg~s$^{-1}$~cm$^{-2}$]&$1.38_{-0.01}^{+0.02}$&$0.95_{-0.02}^{+0.03}$\\
\hline
$\rm plabs^{**}$&$\Delta \Gamma$[NICER/ME]&$0.04_{-0.003}^{+0.005}$&$0.04_{-0.009}^{+0.014}$\\
&$\Delta \Gamma$[NICER/HE]&$0.10_{-0.006}^{+0.004}$&$0.12_{-0.006}^{+0.008}$\\
&$\Delta \Gamma$[NICER/NUSTAR]&$0.12_{-0.005}^{+0.002}$&$0.12_{-0.003}^{+0.005}$\\
&coef[ME]&$0.93_{-0.013}^{+0.013}$&$0.93_{-0.010}^{+0.011}$\\
&coeff[HE]&$1.17_{-0.12}^{+0.13}$&$1.03_{-0.11}^{+0.12}$\\
&coef[NFPMA]&$1.03_{-0.005}^{+0.006}$&$1.03_{-0.005}^{+0.005}$\\
&coef[NFPMB]&$0.71_{-0.003}^{+0.004}$&$0.71_{-0.003}^{+0.003}$\\
\hline
&$\chi^2$/(d.o.f.)&0.86&0.89\\
\hline
	\label{parameter} &
    \end{tabular}
\begin{tablenotes}[para,flushleft] 
        \item Note:\\
        ***We fixed the $\Gamma$ of NICER to 0 and used the $\Delta \Gamma$  to account for the relative index corrections between NuSTAR and Insight-HXMT ME and HE compared to NICER.
     \end{tablenotes} 
     
\end{threeparttable}} 

\end{table}

\begin{table}[htbp]
    \centering
    \renewcommand{\arraystretch}{1.5}
     \setlength{\tabcolsep}{5pt}
		\caption{The results of spectral fitting the NICER+NuSTAR+Insight-HXMT data for Model M2 and M3.}
		\resizebox{8cm}{!}{
		\begin{threeparttable}
	\begin{tabular}{cccc}
		%\begin{tabular}{|c|m{2cm}|}
		
		  \hline
		   \hline
        Model & Parameter &M2&M3
       \\ \hline
tbabs& $N_{\rm H}[10^{22} \rm cm^{-2}]$&$0.15^{+0.003}_{-0.018}$&$0.18^{+0.004}_{-0.011}$\\
\hline
diskbb&$T_{\rm in}$&$0.45_{-0.04}^{+0.01}$&$0.44_{-0.01}^{+0.0016}$ \\
&norm[$10^{4}$]&$6.11_{-0.44}^{+0.55}$&$6.73_{-0.014}^{+0.78}$ \\
\hline
relxill&$a$&$0.98_{-0.08}^{+0.001}$&$0.97_{-0.06}^{+0.02}$ \\
(relxillcp)&$i$ [\textdegree]&$40.5_{-1.02}^{+0.87}$&$39.63_{-0.3}^{+1.36}$\\
&$\Gamma$&$1.23_{-0.02}^{+0.01}$&$1.81_{-0.02}^{+0.01}$ \\
&logxi&$3.04_{-0.32}^{+0.29}$&$2.96_{-0.09}^{+0.06}$\\
&$A_{fe}$&$2.05_{-0.21}^{+0.52}$&$1.31_{-0.32}^{+0.98}$\\
&$E_{\rm cut}$($KT_{\rm e}$)[keV]&$15.09_{-0.17}^{+0.28}$&$32.03_{-2.34}^{+0.15}$\\
&ref\underline{ }frac&$0.18_{-0.02}^{+0.01}$&$0.65_{-0.12}^{+0.02}$\\
&norm&$0.36_{-0.02}^{+0.01}$&$0.16_{-0.003}^{+0.02}$\\
&flux[$10^{-7}$~erg~s$^{-1}$~cm$^{-2}$]&$2.17_{-0.04}^{+0.04}$&$1.27_{-0.03}^{+0.02}$\\
\hline
compTT&$T_{0}$[keV]&$0.23_{-0.02}^{+0.03}$\\
&KT[keV]&$24.74_{-0.31}^{+0.47}$\\
&$\tau$&$1.68_{-0.06}^{+0.08}$\\
&norm&$0.71_{-0.04}^{+0.03}$\\
&flux[$10^{-7}$~erg~s$^{-1}$~cm$^{-2}$]&$0.54_{-0.04}^{+0.04}$ \\
\hline
cutoffpl&PhoIndex&&$1.19_{-0.01}^{+0.02}$\\
&High$E_{\rm cut}$[keV]&&$13.20_{-0.23}^{+0.03}$\\
&norm&&$18.03_{-0.38}^{+0.99}$\\
&flux[$10^{-7}$~erg~s$^{-1}$~cm$^{-2}$]&&$1.54_{-0.03}^{+0.03}$\\
\hline
$\rm plabs^{**}$&$\Delta \Gamma$[NICER/ME]&$0.04_{-0.003}^{+0.005}$&$0.04_{-0.004}^{+0.009}$\\
&$\Delta \Gamma$[NICER/HE]&$0.09_{-0.006}^{+0.004}$&$0.10_{-0.004}^{+0.03}$\\
&$\Delta \Gamma$[NICER/NUSTAR]&$0.12_{-0.005}^{+0.002}$&$0.12_{-0.008}^{+0.001}$\\
&coef[ME]&$0.84_{-0.003}^{+0.003}$&$0.90_{-0.011}^{+0.009}$\\
&coeff[HE]&$1.21_{-0.13}^{+0.13}$&$1.23_{-0.14}^{+0.15}$\\
&coef[NFPMA]&$1.00_{-0.004}^{+0.003}$&$1.00_{-0.004}^{+0.005}$\\
&coef[NFPMB]&$0.70_{-0.003}^{+0.003}$&$0.70_{-0.003}^{+0.003}$\\
\hline
&$\chi^2$/(d.o.f.)&0.87&0.87\\
\hline
	\label{M2} &
    \end{tabular}
\begin{tablenotes}[para,flushleft] 
        \item Note:\\
        ***We fixed the $\Gamma$ of NICER to 0 and used the $\Delta \Gamma$  to account for the relative index corrections between NuSTAR and Insight-HXMT ME and HE compared to NICER.
     \end{tablenotes} 
     
\end{threeparttable}} 

\end{table}

We conduct an analysis of the spectrum of Swift J1727.8--1613 using the FPMA and FPMB data from NuSTAR in the energy range of 3--78 keV.
The first trial for the spectral model is constant*TBABS*(diskbb+relxill), where TBABS accounts for the interstellar absorption \citep{2000Wilms} by considering photoelectric cross-sections provided by \cite{1996Verner}. 
We add the relativistic reflection model RELXILL to fit the reflection component in the spectrum \citep{2016Dauser}.
During fittings,  we fix the two emissivity indices at 3, and the inner radius of the disk at the innermost stable circular orbit (ISCO). At this time, there is a hint for having a hard tail in the residual, with $\chi^2$/(d.o.f.)=1.14 (See figure \ref{residual} (a)).
We also try to replace relxill with relxillcp. The difference between relxillcp and relxill is that they have different incidence spectra, with the former being an nthcomp Comptonization continuum and the latter being a standard high-energy cutoff powerlaw.
However, the fit for the model of the latter is not satisfactory, with $\chi^2$/(d.o.f.)=1.21 (The residuals are shown in Figure \ref{residual} (b)).
In order to better investigate the ``hard tail" and improve the fit by including the simultaneous observations from Insight-HXMT ME and HE, as the maximum energy of NuSTAR is limited to 78 keV. Again, as shown in Figure \ref{residual} (c), the reduced $\chi^2$/(d.o.f.) is as large as 1.37, and a significant residual is shown up at the high energy band.  
To account for the structure observed in the high-energy bands of the residuals, we add a cutoffpl component to the model. Additionally, we used plabs instead of constants to correct for the differences in spectral index between NICER, NuSTAR and Insight-HXMT \citep{2021Zdziarski}, We set K and $\Delta \Gamma$ fixed at 1 and 0, respectively, for the NICER and have a fit of  $\chi^2$/(d.o.f.)=1.02 (See figure \ref{residual} (d)), with a spectral model of  plabs*tbabs(diskbb+relxill+cutoffpl).
To investigate the properties of the disk, we add simultaneous observation from NICER for joint fitting and also add diskbb to the model to account for the multi-temperature blackbody of the accretion disk \citep{1984Mitsuda}, and model 1 adopted is  plabs*tbabs(diskbb+relxill+cutoffpl), $\chi^2$/(d.o.f.)=2601.42/3015=0.86.
In order to investigate whether model plabs may have an impact on our fitting results, we perform a series of trials. First, We replace the plabs with constants and allow only a normalization constant to float between the spectra. Second, We remove the constants, but allow each individual component normalization to vary. Third, we allow each component's normalization and shape parameters to vary.
We find that the normalization and shape parameters are allowed to vary for all components, and the results are essentially the same as those we fitted with plabs. So the extra component can be regarded as robust.

Since in model 1, both the incident spectrum of the reflection and the additional component are cutoff power law, we find that an exchange of the spectral index and cutoff energy between these two components does not change the fitting goodness but however results in completely different descriptions upon the additional spectral component. As shown in Figure \ref{fit}, we find that when the $\Gamma$ and the $E_{\rm cut}$ of the reflected component are larger, the extra component shows up at the mediate energy band as a rather broad hump.  
The count rates of FPMA and FPMB in NuSTAR exhibit differences, but they do not affect the spectral shape.
On the contrary, when the $\Gamma$ and the $E_{\rm cut}$ of the reflected component are set to small values,  the extra component takes a  ``hard tail" located at the high-energy band.
To reduce such a degeneracy in the two spectral components, we replace cutoffpl in model 1 with comptt with geometry as a disk, so model 2 is plabs*tbabs(diskbb+relxill+comptt), and results in  $\chi^2$/(d.o.f.)=2620.02/3014=0.87.  As shown in the left panel of  Figure \ref{re}, in this case, the extra component shows up as  ``hard tail" at the high-energy band.
We also find that the relxill in model 1 can be replaced with  relxillcp to form model 3: plabs*tbabs(diskbb+relxillcp+cutoffpl), which can result in an equally good fit with  $\chi^2$/(d.o.f.)=0.87.
As shown in the right panel of  Figure \ref{re}, the extra component again appears as a broad hump in the mediate energy band.

The parameters from the joint spectral fitting with different models are shown in Tables \ref{parameter} and \ref{M2}, 
where Swift J1727.8--1613 is revealed as a high-spin, medium-inclination system: the spin and the inclination angle are measured as $0.98^{+0.02}_{-0.07}$  and $40^{+1.2}_{-0.8}$ degrees, respectively. 
When a component in addition to the reflection component can either behave as a high energy  ``hard tail", with the flux of the reflection component significantly higher than the flux of the ``hard tail", or a broad hump at the mediate energy band, with its flux comparable to the flux of the reflection component.

\section{discussion and conclusion}
\label{dis}

\begin{table}[htbp]
    \centering
    \renewcommand{\arraystretch}{1.5}
     \setlength{\tabcolsep}{5pt}
		\caption{The results of spectral fitting the NICER+NuSTAR+Insight-HXMT data for Model M4, M5 and M6.}
		\resizebox{8cm}{!}{
		\begin{threeparttable}
	\begin{tabular}{ccccc}
		%\begin{tabular}{|c|m{2cm}|}
		
		  \hline
		   \hline
        Model & Parameter &M4&M5&M6
       \\ \hline
tbabs& $N_{\rm H}[10^{22} \rm cm^{-2}]$&0.15&0.18&0.18\\
\hline
diskbb&$T_{\rm in}$&0.45&0.44&0.45 \\
&norm[$10^{4}$]&6.11&6.73&6.33\\
\hline
relxill&$a$&0.98&0.97&0.98 \\
(relxillcp)&$i$ [\textdegree]&40.5&39.63&41\\
&$\Gamma$&1.23&1.81&1.72 \\
&logxi&3.04&2.96&3.16\\
&$A_{fe}$&2.05&1.31&1.77\\
&$E_{\rm cut}$($KT_{\rm e}$)[keV]&15.09&32.03&59.8\\
&ref\underline{ }frac&0.18&0.6&0.36\\
&norm&0.36&0.16&0.28\\
(relxilllp)&$\beta$[c]&&&$0.15_{-0.04}^{+0.04}$\\
&h[$R_{\rm g}$]&&&$9.01_{-1.0}^{+1.1}$\\
\hline
$\rm relxilllpcp^{*}$ &$\beta$[c]&$0.79_{-0.1}^{+0.1}$&0&\\
($\rm relxilllp^{*}$)&h[$R_{\rm g}$]&500&5&\\
&$\Gamma$&$1.82_{-0.01}^{+0.01}$&$1.21_{-0.01}^{+0.01}$&\\
&$KT_{\rm e}$($E_{\rm cut}$)[keV]&$19.23_{-0.5}^{+0.7}$&$13.4_{-0.5}^{+0.7}$\\
&ref\underline{ }frac&0.2&$0.006_{-0.010}^{+0.005}$ \\
&norm&$0.03_{-0.02}^{+0.01}$&$0.52_{-0.004}^{+0.002}$ \\
\hline
cutoffpl&PhoIndex&&&1.02\\
&High$E_{\rm cut}$[keV]&&&11.45\\
&norm&&&10.53\\
\hline
$\rm plabs^{**}$&$\Delta \Gamma$[NICER/ME]&$0.05_{-0.004}^{+0.005}$&$0.04_{-0.004}^{+0.004}$&$0.04_{-0.004}^{+0.005}$\\
&$\Delta \Gamma$[NICER/HE]&$0.09_{-0.004}^{+0.005}$&$0.11_{-0.004}^{+0.03}$&$0.14_{-0.005}^{+0.004}$\\
&$\Delta \Gamma$[NICER/NUSTAR]&$0.13_{-0.005}^{+0.002}$&$0.11_{-0.003}^{+0.001}$&$0.12_{-0.003}^{+0.004}$\\
&coef[ME]&$0.92_{-0.005}^{+0.004}$&$0.93_{-0.005}^{+0.005}$&$0.93_{-0.005}^{+0.005}$\\
&coeff[HE]&$1.19_{-0.10}^{+0.11}$&$1.20_{-0.10}^{+0.09}$&$1.20_{-0.10}^{+0.10}$\\
&coef[NFPMA]&$1.00_{-0.004}^{+0.003}$&$1.00_{-0.004}^{+0.004}$&$1.00_{-0.004}^{+0.004}$\\
&coef[NFPMB]&$0.71_{-0.003}^{+0.003}$&$0.70_{-0.003}^{+0.003}$&$0.70_{-0.003}^{+0.003}$\\
\hline
&$\chi^2$/(d.o.f.)&0.84&0.84&0.84\\
\hline
	\label{lp} &
    \end{tabular}
\begin{tablenotes}[para,flushleft] 
        \item Note: All parameters that do not have associated errors are fixed based on the previous fitting results for models M2, M3 and M1.\\
        **Same parameters as relxill (relxillcp) we link and fix them.\\
        ***We fixed the $\Gamma$ of NICER to 0 and used the $\Delta \Gamma$  to account for the relative index corrections between NuSTAR and Insight-HXMT ME and HE compared to NICER.
     \end{tablenotes} 
     
\end{threeparttable}} 

\end{table}

\begin{figure*}
\centering

\begin{minipage}[t]{0.45\linewidth}
\centering
\includegraphics[angle=0,scale=0.35]{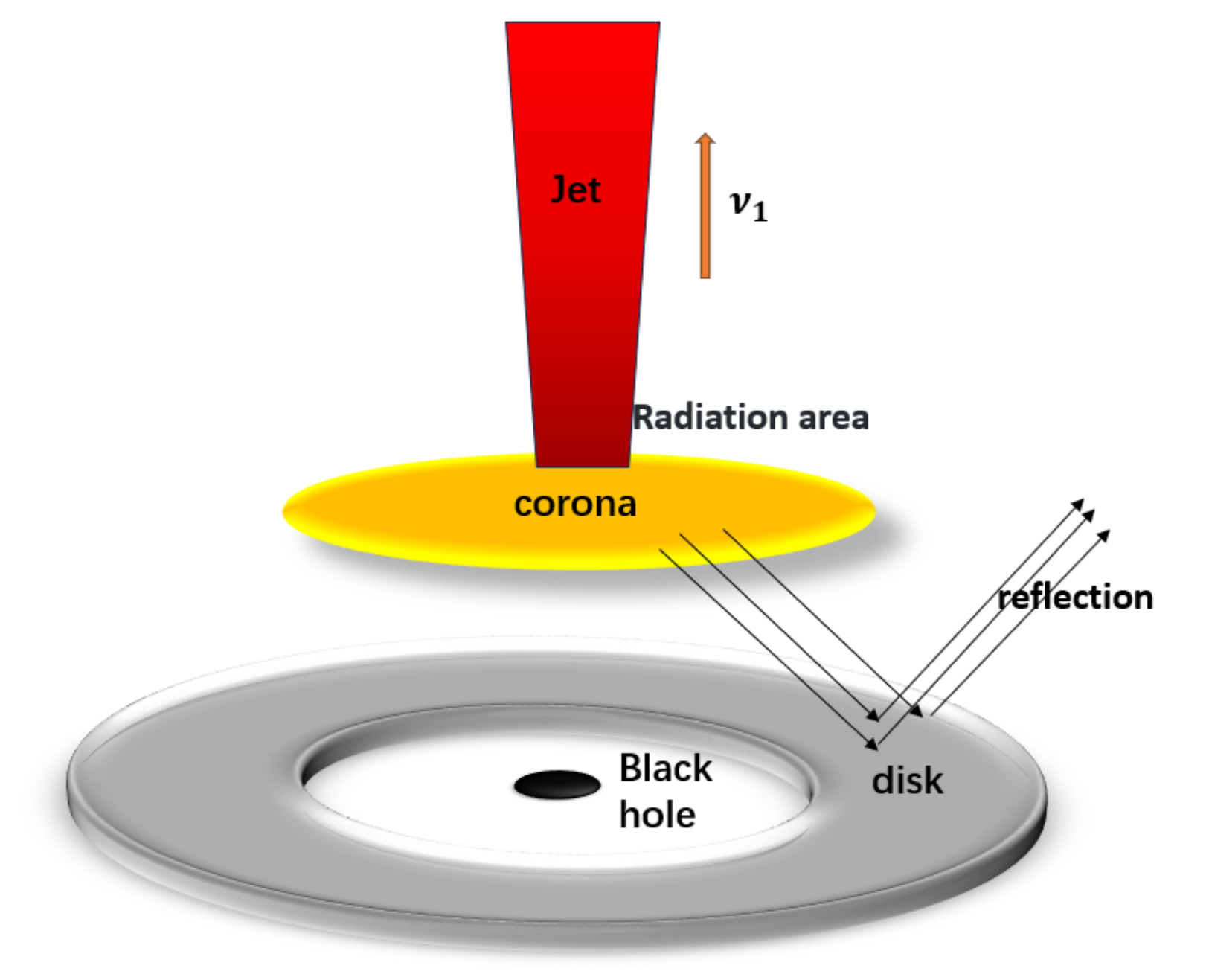}
%\caption{fig1}
\end{minipage}%
\hfill
\begin{minipage}[t]{0.45\linewidth}
\centering
\includegraphics[angle=0,scale=0.35]{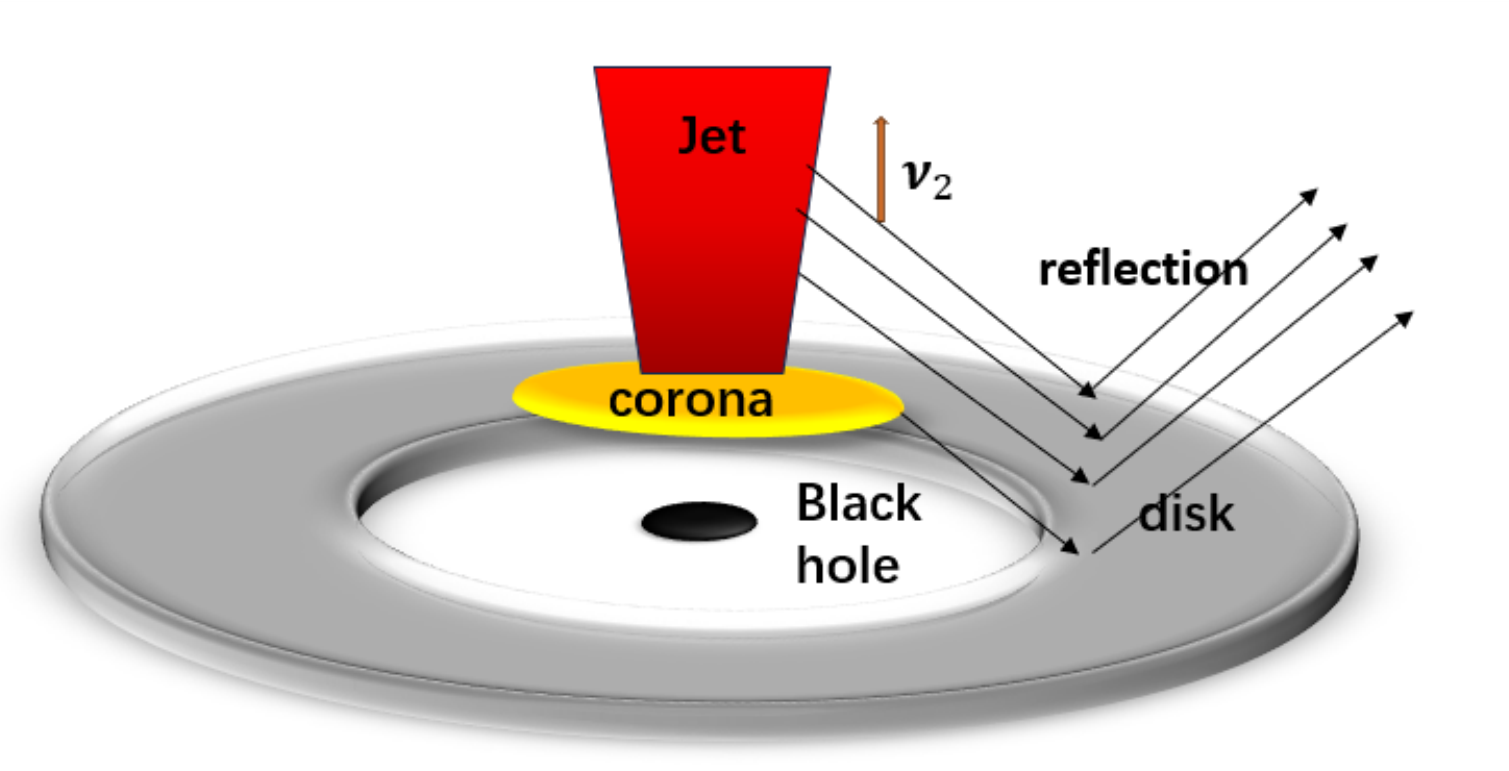}
%\caption{fig2}
\end{minipage}
\centering
\caption{The schematic of the corona/jet of Swift J1727.8--1613. Left panel: the jet has high speed and may as well be blocked by the underneath extended corona, and thus provides less disk reflection.  This configuration corresponds to the fitting results for models M$1^{*}$ and M2,  as shown in the right panel of  Figure \ref{fit} and the left panel of Figure \ref{re}. Right panel: reflection is mainly contributed by the low-speed jet and the underneath jet base/corona is mostly edge-on for the view of the disk and thus provides less reflection than the jet.  This configuration corresponds to the fitting results for models M1 and M3,  as shown in the left panel of  Figure \ref{fit} and the right panel of Figure \ref{re}. }
\label{corona}
\end{figure*}

We have carried out spectral analysis of the 2023 outburst of Swift J1727.8--1613, which was observed simultaneously by Insight-HXMT, NICER and NuSTAR on MJD 60185. 
Such joint observations reveal the properties of the newly discovered BH system Swift J1727.8--1613: the system has a moderate inclination and harbors a highly spinning BH.
This is consistent with the results first reported with NICER data \citep{2023Draghis}. Most interestingly, this system distinguishes from the normal BH X-ray binary system, by showing an additional component in the spectrum apart from the reflection one. 

A spectral analysis solely with NuSTAR observation shows a hint of the need for a hard X-ray tail. Such a component shows up significantly by adding the simultaneous Insight-HXMT observations, which significantly extends the upper bound of the spectrum from  75 keV of NuSTAR to roughly 140 keV.  Accordingly, an additional cutoffpl component is observed apart from the reflection one.  Obviously, such a spectrum is rather peculiar and distinguishable from most of the normal BH X-ray binary systems in their outbursts. 

BH X-ray binary systems usually have a spectrum composed of disk componization and reflection components before entering the soft state of their outbursts but with rare additional non-thermal components e.g. the hard tail.  So far the hard tails have been observed in a few systems. One is from Cyg X--1, mostly in its relatively soft state but not the hard state \citep{2006C,2011L,2014Tomsick}. Systems reminiscent of Swift J1727.8--1613 may be the outbursts from MAXI J1820+070 and MAXI J1535--571.  For MAXI J1820+070, the spectrum observed by Insight-HXMT during the hard state can be well fitted by two reflection components: one from the inner disk and contributing the broadened iron line and the second one illuminates the outer part \citep{2021You},  with both illuminators from the jet.  However, \cite{2023Kawamura} performed a detailed spectral-timing analysis to investigate the nature of the accretion flow in BH 
binaries. By fitting jointly the energy spectrum, the time lag and the QPO rms spectrum, they found that the hard  X-rays at energies above roughly 50 keV can not be well covered by the spectral models, which again may point to the need of a jet emission to account for the possible hard tail. MAXI J1535--571 may be so far the sample most reminiscent of Swift J1727.8--1613: in its hard intermediate state a significant hard tail was observed in addition to the disk reflection component \citep{2020Kong}. 
Obviously, all these hard tails tend to more or less be understood by introducing jet-like structures that produce hard X-rays.  We also notice that \cite{2016Reig} suggested that hard tails may be related to relativistic jets formed by massive collimated magnetic fields near BHs.

Our spectral analysis of Swift J1727.8--1613 suggests two possibilities for describing the extra spectral component. One possibility is that we observed a hard tail at high energies. The presence of an additional hard component, along with the reflection component, in the spectrum of Swift J1727.8--1613 provides an additional sample in such a hard-tail family of BHXRBs. An alternative possibility is that the extra spectral component may take a broad hump shape in the mediate energy band. We constitute in what follows a joint jet-corona scenario and adjust the configuration to account for both cases of the inferred possible extra spectral component.

By introducing Swift J1727.8--1613 a jet-corona scenario, one key point is how to handle the additional component that may contribute to less disk reflections. 
If the extra component is a hard tail, one possibility is that the jet responsible for the hard tail has a large velocity and thus results in less illumination of the disk.
In our fitting (M2), we replace comptt with relxilllpcp to form model 4 (M4) but encounter difficulty in constraining the height and reflection fraction. Consequently, if we fix them at 500 $R_{\rm g}$ and 0.2, respectively, the velocity of the jet is estimated around  $0.79_{-0.1}^{+0.1}$c (see Table \ref{lp}). We notice that having such a large velocity is independent of how to choose the height. As a result, the contribution of the jet producing the hard tail to the reflection component is small (see the left panel of Figure \ref{corona}).  
The reflection component mainly comes from the high jet base/corona, which hangs above the disk. 
The corona has the potential to block the jet,
if the jet base/corona has a relatively large extension in the direction perpendicular to the jet and a large optical depth. The optical depth is estimated as about 3.61 according to the parameter $\Gamma$ and $E_{\rm cut}$ listed in Table \ref{parameter}. Therefore, this blocking may lead to even less jet contribution to the disk reflection.

Spectral fittings show the extra component of a broad hump can also show up in the mediate energy band.
In our investigation of M5, where we replace cutoffpl in M3 with relxilllp and fix its reflection parameters, we find that, while the velocity and height of this additional reflection component can not be constrained, the reflection fraction has a rather small value. As shown in Table \ref{lp}, assuming that the additional reflection component originates from the corona, we fix the velocity at 0 and obtain a reflection fraction of approximately $0.006_{-0.005}^{+0.010}$.
As shown in the right panel of Figure \ref{corona},  the jet base/corona can be located at a site very close to the BH, and the jet on top of the jet base/corona dominates the disk illumination. In such a configuration, a flat jet base/corona located close to the BH can be viewed by the disk from mostly edge-on, and hence contribute reflection less dominant than the jet. 
To investigate the height and velocity of the jet responsible for generating reflections, we replace relxill in M1 with relxilllp to form model 6 (M6) and fix its parameters other than height and velocity. It turns out that the height and velocity can be well measured as  $9.01_{-1.0}^{+1.1} R_{\rm g}$ and $0.15_{-0.04}^{+0.04}$c, respectively (see Table \ref{lp}). The corona may resist gravity through a magnetic field to allow it to exist at several times $R_{g}$ above the BH. \cite{2023Y} discovered that the presence of a sufficiently strong magnetic field in the radial direction of the disk can halt the accretion flow, producing a magnetically arrested disk (MAD) in MAXI J1820+070. The strength of the magnetic field in both the vertical and radial directions is typically considered to be of the same order of magnitude, and we hypothesize the existence of an extremely strong magnetic field in the vertical direction to counteract gravity.

These results support further the scenario of having a low speed jet and an underneath jet base/corona located close to the BH, as shown in the right panel of Figure \ref{corona}. 

Another issue is how the hard component without obvious reflection is produced. It is commonly believed that hard X-ray photons from BH X-ray binaries are produced by inverse Compton scattering of low energy photons (also called seed photons) by hot or high energy electrons, as the comptt model describes. The seed photons are commonly believed to be produced by the thermal emission from the disk. However, the lack of reflection means that the hard X-ray photons do not illuminate the disk effectively, and consequently the disk photons also cannot reach the jet/corona producing the hard photons. It seems the only option out is that the seed photons are produced inside the jet/corona, perhaps through synchrotron radiation of high energy electrons in strong magnetic fields. This is similar to the so-called synchrotron-self Comptonization process commonly used in describing the high energy spectra of the jets of Blazars and gamma-ray bursts. Further theoretical studies are needed to understand this problem.

In summary, simultaneous observations from Insight-HXMT, NICER and NuSTAR provide the first view of the possible intrinsic properties of the newly discovered BH X-ray binary system Swift J1727.8--1613. Apart from revealing the spin of the BH and the inclination of the system, the additional non-thermal spectral component discovered for Swift J1727.8--1613 may need a jet-corona configuration to account for the results from different trails of spectral fittings. 

%% Also note that the akcnowlodgment environment does not support long amounts of text. If you have a lot of people and institutions to acknowledge, do not use this command. Instead, create a new \section{Acknowledgments}.
\begin{acknowledgments}
This work is supported by the National Key R\&D Program of China (2021YFA0718500), the National Natural Science Foundation of China under grants U1838201, U1838202,  U2038101, U1938103, 12273030, U1938107.
This work made use of data and software from the \textit{Insight}-HXMT mission, a project funded by the China National Space Administration (CNSA) and the Chinese Academy of Sciences(CAS). This work was partially supported by the International Partnership Program of the Chinese Academy of Sciences (Grant No.113111KYSB20190020).
This research has made use of software provided by data obtained from the High Energy Astrophysics Science Archive Research Center (HEASARC), provided by NASA’s Goddard Space Flight Center.
L. D. Kong is grateful for the financial support provided by the Sino-German (CSC-DAAD) Postdoc Scholarship Program (57251553).
\end{acknowledgments}

\appendix

\begin{figure}
	\centering
	\includegraphics[angle=0,scale=0.2]{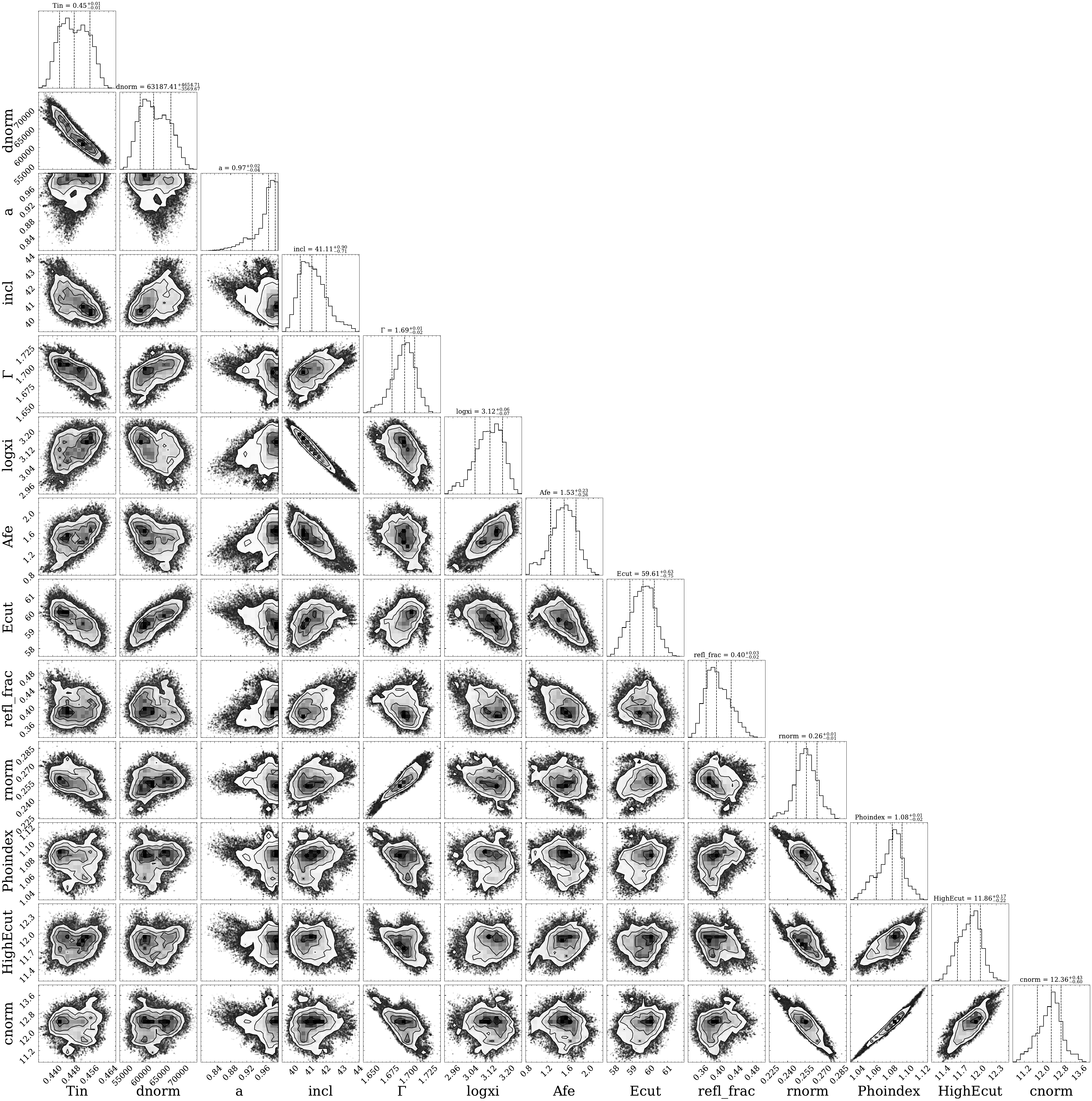}
	\caption{An illustration of one and two-dimensional projections of the posterior probability distributions derived from the MCMC analysis for the parameters in M1  }
	\label{con}
\end{figure}

\begin{figure}
	\centering
	\includegraphics[angle=0,scale=0.2]{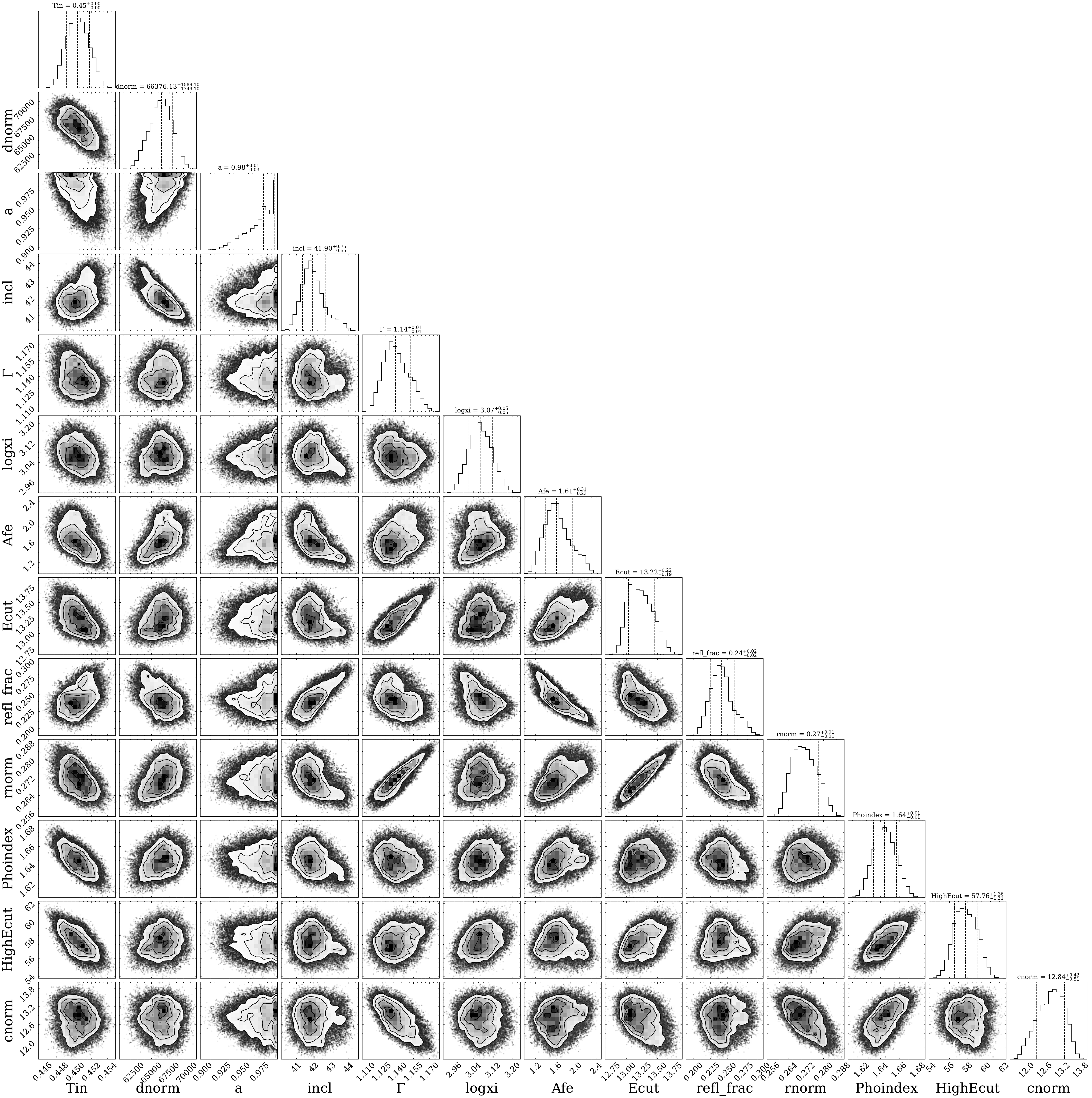}
	\caption{An illustration of one and two-dimensional projections of the posterior probability distributions derived from the MCMC analysis for the parameters in M$1^*$  }
	\label{con}
\end{figure}






\bibliography{ref}{}

\begin{thebibliography}{}
\expandafter\ifx\csname natexlab\endcsname\relax\def\natexlab#1{#1}\fi
\providecommand{\url}[1]{\href{#1}{#1}}
\providecommand{\dodoi}[1]{doi:~\href{http://doi.org/#1}{\nolinkurl{#1}}}
\providecommand{\doeprint}[1]{\href{http://ascl.net/#1}{\nolinkurl{http://ascl.net/#1}}}
\providecommand{\doarXiv}[1]{\href{https://arxiv.org/abs/#1}{\nolinkurl{https://arxiv.org/abs/#1}}}

\bibitem[{{Belloni} {et~al.}(2005){Belloni}, {Homan}, {Casella}, {van der Klis}, {Nespoli}, {Lewin}, {Miller}, \& {M{\'e}ndez}}]{2005Belloni}
{Belloni}, T., {Homan}, J., {Casella}, P., {et~al.} 2005, \aap, 440, 207, \dodoi{10.1051/0004-6361:20042457}

\bibitem[{{Belloni} {et~al.}(2011){Belloni}, {Motta}, \& {Mu{\~n}oz-Darias}}]{2011Belloni}
{Belloni}, T.~M., {Motta}, S.~E., \& {Mu{\~n}oz-Darias}, T. 2011, Bulletin of the Astronomical Society of India, 39, 409, \dodoi{10.48550/arXiv.1109.3388}

\bibitem[{{Brenneman} \& {Reynolds}(2006)}]{2006Brenneman}
{Brenneman}, L.~W., \& {Reynolds}, C.~S. 2006, \apj, 652, 1028, \dodoi{10.1086/508146}

\bibitem[{{Cadolle Bel} {et~al.}(2006){Cadolle Bel}, {Sizun}, {Goldwurm}, {Rodriguez}, {Laurent}, {Zdziarski}, {Foschini}, {Goldoni}, {Gouiff{\`e}s}, {Malzac}, {Jourdain}, \& {Roques}}]{2006C}
{Cadolle Bel}, M., {Sizun}, P., {Goldwurm}, A., {et~al.} 2006, \aap, 446, 591, \dodoi{10.1051/0004-6361:20053068}

\bibitem[{{Cannizzo} {et~al.}(1995){Cannizzo}, {Chen}, \& {Livio}}]{1995Cannizzo}
{Cannizzo}, J.~K., {Chen}, W., \& {Livio}, M. 1995, \apj, 454, 880, \dodoi{10.1086/176541}

\bibitem[{{Cao} {et~al.}(2020){Cao}, {Jiang}, {Meng}, {Zhang}, {Luo}, {Yang}, {Zhang}, {Gu}, {Sun}, {Liu}, {Yang}, {Li}, {Tan}, {Liu}, {Du}, {Lu}, {Xu}, {Guan}, {Zhang}, {Wang}, {Li}, {Zhang}, {Wen}, {Qu}, {Song}, {Li}, {Ge}, {Zhou}, {Xiong}, {Zhang}, {Zhang}, {Cheng}, {Zhang}, {Li}, {Liang}, {Gao}, {Yang}, {Liu}, {Liu}, {Yang}, \& {Zhang}}]{2020Cao}
{Cao}, X., {Jiang}, W., {Meng}, B., {et~al.} 2020, Science China Physics, Mechanics, and Astronomy, 63, 249504, \dodoi{10.1007/s11433-019-1506-1}

\bibitem[{{Castro-Tirado} {et~al.}(2023){Castro-Tirado}, {Sanchez-Ramirez}, {Caballero-Garcia}, {Perez-Garcia}, {Fernandez-Garcia}, {Guziy}, {Hu}, {Blazek}, {Hermelo}, {Pinter}, {Meintjes}, {van Heerden}, {Martin-Carrillo}, {Hanlon}, {Hiriart}, {Lee}, {Carrasco-Garcia}, {Park}, {Gritsevich}, {Castellon}, {Perez del Pulgar}, \& {Reina}}]{2023Castro-Tirado}
{Castro-Tirado}, A.~J., {Sanchez-Ramirez}, R., {Caballero-Garcia}, M.~D., {et~al.} 2023, The Astronomer's Telegram, 16208, 1

\bibitem[{{Chen} {et~al.}(2020){Chen}, {Cui}, {Li}, {Wang}, {Xu}, {Lu}, {Wang}, {Chen}, {Han}, {Hu}, {Zhang}, {Huo}, {Yang}, {Li}, {Lu}, {Zhang}, {Li}, {Zhang}, {Xiong}, {Zhang}, {Xue}, {Zhao}, {Zhu}, {Zhu}, {Liu}, {Yang}, \& {Zhang}}]{2020Chen}
{Chen}, Y., {Cui}, W., {Li}, W., {et~al.} 2020, Science China Physics, Mechanics, and Astronomy, 63, 249505, \dodoi{10.1007/s11433-019-1469-5}

\bibitem[{{Corral-Santana} {et~al.}(2016){Corral-Santana}, {Casares}, {Mu{\~n}oz-Darias}, {Bauer}, {Mart{\'\i}nez-Pais}, \& {Russell}}]{2016Corral-Santana}
{Corral-Santana}, J.~M., {Casares}, J., {Mu{\~n}oz-Darias}, T., {et~al.} 2016, \aap, 587, A61, \dodoi{10.1051/0004-6361/201527130}

\bibitem[{{Dauser} {et~al.}(2016){Dauser}, {Garc{\'\i}a}, \& {Wilms}}]{2016Dauser}
{Dauser}, T., {Garc{\'\i}a}, J., \& {Wilms}, J. 2016, Astronomische Nachrichten, 337, 362, \dodoi{10.1002/asna.201612314}

\bibitem[{{Draghis} {et~al.}(2023){Draghis}, {Miller}, {Homan}, {Uttley}, {Bollemeijer}, {Steiner}, {Hare}, {Tombesi}, {Gendreau}, {Arzoumanian}, {Strohmayer}, {Sanna}, {Altamirano}, {Buisson}, \& {Fabian}}]{2023Draghis}
{Draghis}, P.~A., {Miller}, J.~M., {Homan}, J., {et~al.} 2023, The Astronomer's Telegram, 16219, 1

\bibitem[{{Esin} {et~al.}(1997){Esin}, {McClintock}, \& {Narayan}}]{1997Esin}
{Esin}, A.~A., {McClintock}, J.~E., \& {Narayan}, R. 1997, \apj, 489, 865, \dodoi{10.1086/304829}

\bibitem[{{Gendreau} {et~al.}(2016){Gendreau}, {Arzoumanian}, {Adkins}, {Albert}, {Anders}, {Aylward}, {Baker}, {Balsamo}, {Bamford}, {Benegalrao}, {Berry}, {Bhalwani}, {Black}, {Blaurock}, {Bronke}, {Brown}, {Budinoff}, {Cantwell}, {Cazeau}, {Chen}, {Clement}, {Colangelo}, {Coleman}, {Coopersmith}, {Dehaven}, {Doty}, {Egan}, {Enoto}, {Fan}, {Ferro}, {Foster}, {Galassi}, {Gallo}, {Green}, {Grosh}, {Ha}, {Hasouneh}, {Heefner}, {Hestnes}, {Hoge}, {Jacobs}, {J{\o}rgensen}, {Kaiser}, {Kellogg}, {Kenyon}, {Koenecke}, {Kozon}, {LaMarr}, {Lambertson}, {Larson}, {Lentine}, {Lewis}, {Lilly}, {Liu}, {Malonis}, {Manthripragada}, {Markwardt}, {Matonak}, {Mcginnis}, {Miller}, {Mitchell}, {Mitchell}, {Mohammed}, {Monroe}, {Montt de Garcia}, {Mul{\'e}}, {Nagao}, {Ngo}, {Norris}, {Norwood}, {Novotka}, {Okajima}, {Olsen}, {Onyeachu}, {Orosco}, {Peterson}, {Pevear}, {Pham}, {Pollard}, {Pope}, {Powers}, {Powers}, {Price}, {Prigozhin}, {Ramirez}, {Reid}, {Remillard}, {Rogstad}, {Rosecrans}, {Rowe}, {Sager}, {Sanders},
  {Savadkin}, {Saylor}, {Schaeffer}, {Schweiss}, {Semper}, {Serlemitsos}, {Shackelford}, {Soong}, {Struebel}, {Vezie}, {Villasenor}, {Winternitz}, {Wofford}, {Wright}, {Yang}, \& {Yu}}]{2016Gendreau}
{Gendreau}, K.~C., {Arzoumanian}, Z., {Adkins}, P.~W., {et~al.} 2016, in Society of Photo-Optical Instrumentation Engineers (SPIE) Conference Series, Vol. 9905, Space Telescopes and Instrumentation 2016: Ultraviolet to Gamma Ray, ed. J.-W.~A. {den Herder}, T.~{Takahashi}, \& M.~{Bautz}, 99051H, \dodoi{10.1117/12.2231304}

\bibitem[{{Harrison} {et~al.}(2013){Harrison}, {Craig}, {Christensen}, {Hailey}, {Zhang}, {Boggs}, {Stern}, {Cook}, {Forster}, {Giommi}, {Grefenstette}, {Kim}, {Kitaguchi}, {Koglin}, {Madsen}, {Mao}, {Miyasaka}, {Mori}, {Perri}, {Pivovaroff}, {Puccetti}, {Rana}, {Westergaard}, {Willis}, {Zoglauer}, {An}, {Bachetti}, {Barri{\`e}re}, {Bellm}, {Bhalerao}, {Brejnholt}, {Fuerst}, {Liebe}, {Markwardt}, {Nynka}, {Vogel}, {Walton}, {Wik}, {Alexander}, {Cominsky}, {Hornschemeier}, {Hornstrup}, {Kaspi}, {Madejski}, {Matt}, {Molendi}, {Smith}, {Tomsick}, {Ajello}, {Ballantyne}, {Balokovi{\'c}}, {Barret}, {Bauer}, {Blandford}, {Brandt}, {Brenneman}, {Chiang}, {Chakrabarty}, {Chenevez}, {Comastri}, {Dufour}, {Elvis}, {Fabian}, {Farrah}, {Fryer}, {Gotthelf}, {Grindlay}, {Helfand}, {Krivonos}, {Meier}, {Miller}, {Natalucci}, {Ogle}, {Ofek}, {Ptak}, {Reynolds}, {Rigby}, {Tagliaferri}, {Thorsett}, {Treister}, \& {Urry}}]{2013Harrison}
{Harrison}, F.~A., {Craig}, W.~W., {Christensen}, F.~E., {et~al.} 2013, \apj, 770, 103, \dodoi{10.1088/0004-637X/770/2/103}

\bibitem[{{Homan} \& {Belloni}(2005)}]{2005Homan}
{Homan}, J., \& {Belloni}, T. 2005, \apss, 300, 107, \dodoi{10.1007/s10509-005-1197-4}

\bibitem[{{Kawamura} {et~al.}(2023){Kawamura}, {Done}, {Axelsson}, \& {Takahashi}}]{2023Kawamura}
{Kawamura}, T., {Done}, C., {Axelsson}, M., \& {Takahashi}, T. 2023, \mnras, 519, 4434, \dodoi{10.1093/mnras/stad014}

\bibitem[{{Kong} {et~al.}(2020){Kong}, {Zhang}, {Chen}, {Ji}, {Zhang}, {Yang}, {Tao}, {Ma}, {Qu}, {Lu}, {Bu}, {Chen}, {Song}, {Li}, {Xu}, {Cao}, {Chen}, {Liu}, {Cai}, {Chang}, {Chen}, {Chen}, {Chen}, {Cui}, {Cui}, {Deng}, {Dong}, {Du}, {Fu}, {Gao}, {Gao}, {Gao}, {Ge}, {Gu}, {Guan}, {Guo}, {Han}, {Huang}, {Huo}, {Jia}, {Jiang}, {Jiang}, {Jin}, {Li}, {Li}, {Li}, {Li}, {Li}, {Li}, {Li}, {Li}, {Li}, {Li}, {Liang}, {Liao}, {Liu}, {Liu}, {Liu}, {Liu}, {Liu}, {Liu}, {Lu}, {Lu}, {Luo}, {Luo}, {Meng}, {Nang}, {Nie}, {Ou}, {Ren}, {Sai}, {Song}, {Sun}, {Tan}, {Tuo}, {Wang}, {Wang}, {Wang}, {Wang}, {Wang}, {Wang}, {Wen}, {Wu}, {Wu}, {Wu}, {Xiao}, {Xiao}, {Xiong}, {Xu}, {Yang}, {Yang}, {Yang}, {Yi}, {You}, {Zhang}, {Zhang}, {Zhang}, {Zhang}, {Zhang}, {Zhang}, {Zhang}, {Zhang}, {Zhang}, {Zhang}, {Zhang}, {Zhang}, {Zhang}, {Zhang}, {Zhang}, {Zhang}, {Zhang}, {Zhao}, {Zhao}, {Zheng}, {Zheng}, {Zhou}, {Zhou}, {Zhu}, {Zhu}, \& {Insight-HXMT Collaboration}}]{2020Kong}
{Kong}, L.~D., {Zhang}, S., {Chen}, Y.~P., {et~al.} 2020, Journal of High Energy Astrophysics, 25, 29, \dodoi{10.1016/j.jheap.2020.01.003}

\bibitem[{{Kong} {et~al.}(2021){Kong}, {Zhang}, {Chen}, {Zhang}, {Ji}, {Wang}, {Tao}, {Ge}, {Liu}, {Song}, {Lu}, {Qu}, {Li}, {Xu}, {Cao}, {Chen}, {Bu}, {Cai}, {Chang}, {Chen}, {Chen}, {Chen}, {Cui}, {Du}, {Gao}, {Gao}, {Gao}, {Gu}, {Guan}, {Guo}, {Han}, {Huang}, {Huo}, {Jia}, {Jiang}, {Jin}, {Li}, {Li}, {Li}, {Li}, {Li}, {Li}, {Li}, {Li}, {Liang}, {Liao}, {Liu}, {Liu}, {Liu}, {Liu}, {Lu}, {Luo}, {Luo}, {Ma}, {Ma}, {Meng}, {Nang}, {Nie}, {Ou}, {Ren}, {Sai}, {Song}, {Sun}, {Tan}, {Tuo}, {Wang}, {Wang}, {Wang}, {Wang}, {Wen}, {Wu}, {Wu}, {Wu}, {Xiao}, {Xiao}, {Xiong}, {Yang}, {Yang}, {Yang}, {Yang}, {Yi}, {Yin}, {You}, {Zhang}, {Zhang}, {Zhang}, {Zhang}, {Zhang}, {Zhang}, {Zhang}, {Zhang}, {Zhao}, {Zhao}, {Zheng}, {Zheng}, \& {Zhou}}]{2021Kong}
---. 2021, \apjl, 906, L2, \dodoi{10.3847/2041-8213/abd03d}

\bibitem[{{Lasota}(2001)}]{2001Lasota}
{Lasota}, J.-P. 2001, \nar, 45, 449, \dodoi{10.1016/S1387-6473(01)00112-9}

\bibitem[{{Laurent} {et~al.}(2011){Laurent}, {Rodriguez}, {Wilms}, {Cadolle Bel}, {Pottschmidt}, \& {Grinberg}}]{2011L}
{Laurent}, P., {Rodriguez}, J., {Wilms}, J., {et~al.} 2011, Science, 332, 438, \dodoi{10.1126/science.1200848}

\bibitem[{{Li} {et~al.}(2005){Li}, {Zimmerman}, {Narayan}, \& {McClintock}}]{2005Li}
{Li}, L.-X., {Zimmerman}, E.~R., {Narayan}, R., \& {McClintock}, J.~E. 2005, \apjs, 157, 335, \dodoi{10.1086/428089}

\bibitem[{{Liu} {et~al.}(2020){Liu}, {Zhang}, {Li}, {Lu}, {Chang}, {Li}, {Zhang}, {Jin}, {Yu}, {Zhang}, {Fu}, {Chen}, {Ji}, {Xu}, {Deng}, {Shang}, {Liu}, {Lu}, {Zhang}, {Dong}, {Li}, {Wu}, {Li}, {Wang}, {Wu}, {Zhang}, {Zhang}, {Xiong}, {Liu}, {Zhang}, {Liu}, {Yang}, \& {Zhang}}]{2020Liu}
{Liu}, C., {Zhang}, Y., {Li}, X., {et~al.} 2020, Science China Physics, Mechanics, and Astronomy, 63, 249503, \dodoi{10.1007/s11433-019-1486-x}

\bibitem[{{McClintock} \& {Remillard}(2006)}]{2006McClintock}
{McClintock}, J.~E., \& {Remillard}, R.~A. 2006, in Compact stellar X-ray sources, Vol.~39, 157--213, \dodoi{10.48550/arXiv.astro-ph/0306213}

\bibitem[{{Miller} {et~al.}(2009){Miller}, {Reynolds}, {Fabian}, {Miniutti}, \& {Gallo}}]{2009Miller}
{Miller}, J.~M., {Reynolds}, C.~S., {Fabian}, A.~C., {Miniutti}, G., \& {Gallo}, L.~C. 2009, \apj, 697, 900, \dodoi{10.1088/0004-637X/697/1/900}

\bibitem[{{Miller-Jones} {et~al.}(2023){Miller-Jones}, {Sivakoff}, {Bahramian}, \& {Russell}}]{2023Miller}
{Miller-Jones}, J.~C.~A., {Sivakoff}, G.~R., {Bahramian}, A., \& {Russell}, T.~D. 2023, The Astronomer's Telegram, 16211, 1

\bibitem[{{Mitsuda} {et~al.}(1984){Mitsuda}, {Inoue}, {Koyama}, {Makishima}, {Matsuoka}, {Ogawara}, {Shibazaki}, {Suzuki}, {Tanaka}, \& {Hirano}}]{1984Mitsuda}
{Mitsuda}, K., {Inoue}, H., {Koyama}, K., {et~al.} 1984, \pasj, 36, 741

\bibitem[{{Motta} {et~al.}(2009){Motta}, {Belloni}, \& {Homan}}]{2009Motta}
{Motta}, S., {Belloni}, T., \& {Homan}, J. 2009, \mnras, 400, 1603, \dodoi{10.1111/j.1365-2966.2009.15566.x}

\bibitem[{{Nakajima} {et~al.}(2023){Nakajima}, {Negoro}, {Serino}, {Mihara}, {Kobayashi}, {Tanaka}, {Soejima}, {Kudo}, {Kawamuro}, {Yamada}, {Tamagawa}, {Kawai}, {Matsuoka}, {Sakamoto}, {Sugita}, {Hiramatsu}, {Nishikawa}, {Yoshida}, {Tsuboi}, {Urabe}, {Nawa}, {Nemoto}, {Shidatsu}, {Takahashi}, {Niwano}, {Sato}, {Higuchi}, {Yatsu}, {Nakahira}, {Ueno}, {Tomida}, {Ishikawa}, {Ogawa}, {Kurihara}, {Ueda}, {Setoguchi}, {Yoshitake}, {Nakatani}, {Yamauchi}, {Hagiwara}, {Umeki}, {Otsuki}, {Yamaoka}, {Kawakubo}, {Sugizaki}, \& {Iwakiri}}]{2023Nakajima}
{Nakajima}, M., {Negoro}, H., {Serino}, M., {et~al.} 2023, The Astronomer's Telegram, 16206, 1

\bibitem[{{O'Connor} {et~al.}(2023){O'Connor}, {Hare}, {Younes}, {Gendreau}, {Arzoumanian}, \& {Ferrara}}]{2023O'Connor}
{O'Connor}, B., {Hare}, J., {Younes}, G., {et~al.} 2023, The Astronomer's Telegram, 16207, 1

\bibitem[{{Palmer} \& {Parsotan}(2023)}]{2023Palmer}
{Palmer}, D.~M., \& {Parsotan}, T.~M. 2023, The Astronomer's Telegram, 16215, 1

\bibitem[{{Reig} \& {Kylafis}(2016)}]{2016Reig}
{Reig}, P., \& {Kylafis}, N. 2016, \aap, 591, A24, \dodoi{10.1051/0004-6361/201628294}

\bibitem[{{Shakura} \& {Sunyaev}(1973)}]{1973Shakura}
{Shakura}, N.~I., \& {Sunyaev}, R.~A. 1973, \aap, 24, 337

\bibitem[{{Sreehari} {et~al.}(2018){Sreehari}, {Nandi}, {Radhika}, {Iyer}, \& {Mandal}}]{2018Sreehari}
{Sreehari}, H., {Nandi}, A., {Radhika}, D., {Iyer}, N., \& {Mandal}, S. 2018, Journal of Astrophysics and Astronomy, 39, 5, \dodoi{10.1007/s12036-018-9510-0}

\bibitem[{{Tetarenko} {et~al.}(2016){Tetarenko}, {Sivakoff}, {Heinke}, \& {Gladstone}}]{2016Tetarenko}
{Tetarenko}, B.~E., {Sivakoff}, G.~R., {Heinke}, C.~O., \& {Gladstone}, J.~C. 2016, VizieR Online Data Catalog, J/ApJS/222/15

\bibitem[{{Tomsick} {et~al.}(2014){Tomsick}, {Nowak}, {Parker}, {Miller}, {Fabian}, {Harrison}, {Bachetti}, {Barret}, {Boggs}, {Christensen}, {Craig}, {Forster}, {F{\"u}rst}, {Grefenstette}, {Hailey}, {King}, {Madsen}, {Natalucci}, {Pottschmidt}, {Ross}, {Stern}, {Walton}, {Wilms}, \& {Zhang}}]{2014Tomsick}
{Tomsick}, J.~A., {Nowak}, M.~A., {Parker}, M., {et~al.} 2014, \apj, 780, 78, \dodoi{10.1088/0004-637X/780/1/78}

\bibitem[{{Verner} {et~al.}(1996){Verner}, {Ferland}, {Korista}, \& {Yakovlev}}]{1996Verner}
{Verner}, D.~A., {Ferland}, G.~J., {Korista}, K.~T., \& {Yakovlev}, D.~G. 1996, \apj, 465, 487, \dodoi{10.1086/177435}

\bibitem[{{Wilms} {et~al.}(2000){Wilms}, {Allen}, \& {McCray}}]{2000Wilms}
{Wilms}, J., {Allen}, A., \& {McCray}, R. 2000, \apj, 542, 914, \dodoi{10.1086/317016}

\bibitem[{{You} {et~al.}(2021){You}, {Tuo}, {Li}, {Wang}, {Zhang}, {Zhang}, {Ge}, {Luo}, {Liu}, {Yuan}, {Dai}, {Liu}, {Qiao}, {Jin}, {Liu}, {Czerny}, {Wu}, {Bu}, {Cai}, {Cao}, {Chang}, {Chen}, {Chen}, {Chen}, {Chen}, {Chen}, {Chen}, {Cui}, {Cui}, {Deng}, {Dong}, {Du}, {Fu}, {Gao}, {Gao}, {Gao}, {Gu}, {Guan}, {Guo}, {Han}, {Huang}, {Huo}, {Jia}, {Jiang}, {Jiang}, {Jin}, {Jin}, {Kong}, {Li}, {Li}, {Li}, {Li}, {Li}, {Li}, {Li}, {Li}, {Li}, {Li}, {Li}, {Liang}, {Liao}, {Liu}, {Liu}, {Liu}, {Liu}, {Liu}, {Lu}, {Lu}, {Lu}, {Luo}, {Luo}, {Ma}, {Meng}, {Nang}, {Nie}, {Ou}, {Qu}, {Sai}, {Shang}, {Song}, {Song}, {Sun}, {Tan}, {Tao}, {Wang}, {Wang}, {Wang}, {Wang}, {Wang}, {Wang}, {Wen}, {Wu}, {Wu}, {Wu}, {Xiao}, {Xiao}, {Xiong}, {Xu}, {Yang}, {Yang}, {Yang}, {Yi}, {Yin}, {You}, {Zhang}, {Zhang}, {Zhang}, {Zhang}, {Zhang}, {Zhang}, {Zhang}, {Zhang}, {Zhang}, {Zhang}, {Zhang}, {Zhang}, {Zhang}, {Zhang}, {Zhang}, {Zhao}, {Zhao}, {Zheng}, {Zhou}, {Zhou}, {Zhu}, \& {Zhu}}]{2021You}
{You}, B., {Tuo}, Y., {Li}, C., {et~al.} 2021, Nature Communications, 12, 1025, \dodoi{10.1038/s41467-021-21169-5}

\bibitem[{{You} {et~al.}(2023){You}, {Cao}, {Yan}, {Hameury}, {Czerny}, {Wu}, {Xia}, {Sikora}, {Zhang}, {Du}, \& {Zycki}}]{2023Y}
{You}, B., {Cao}, X., {Yan}, Z., {et~al.} 2023, Science, 381, 961, \dodoi{10.1126/science.abo4504}

\bibitem[{{Zdziarski} {et~al.}(2021){Zdziarski}, {Jourdain}, {Lubi{\'n}ski}, {Szanecki}, {Nied{\'z}wiecki}, {Veledina}, {Poutanen}, {Dzie{\l}ak}, \& {Roques}}]{2021Zdziarski}
{Zdziarski}, A.~A., {Jourdain}, E., {Lubi{\'n}ski}, P., {et~al.} 2021, \apjl, 914, L5, \dodoi{10.3847/2041-8213/ac0147}

\bibitem[{{Zhang} {et~al.}(2014){Zhang}, {Lu}, {Zhang}, \& {Li}}]{2014Zhang}
{Zhang}, S., {Lu}, F.~J., {Zhang}, S.~N., \& {Li}, T.~P. 2014, in Society of Photo-Optical Instrumentation Engineers (SPIE) Conference Series, Vol. 9144, Space Telescopes and Instrumentation 2014: Ultraviolet to Gamma Ray, ed. T.~{Takahashi}, J.-W.~A. {den Herder}, \& M.~{Bautz}, 914421, \dodoi{10.1117/12.2054144}

\bibitem[{{Zhang} {et~al.}(2018){Zhang}, {Zhang}, {Lu}, {Li}, {Song}, {Xu}, {Wang}, {Qu}, {Liu}, {Chen}, {Cao}, {Zhang}, {Xiong}, {Ge}, {Chen}, {Liao}, {Nie}, {Zhao}, {Jia}, {Li}, {Guan}, {Li}, {Zhang}, {Jin}, {Wang}, {Zheng}, {Ma}, {Tao}, \& {Huang}}]{2018Zhang}
{Zhang}, S., {Zhang}, S.~N., {Lu}, F.~J., {et~al.} 2018, in Society of Photo-Optical Instrumentation Engineers (SPIE) Conference Series, Vol. 10699, Space Telescopes and Instrumentation 2018: Ultraviolet to Gamma Ray, ed. J.-W.~A. {den Herder}, S.~{Nikzad}, \& K.~{Nakazawa}, 106991U, \dodoi{10.1117/12.2311835}

\bibitem[{{Zhang} {et~al.}(1997){Zhang}, {Cui}, \& {Chen}}]{1997Zhang}
{Zhang}, S.~N., {Cui}, W., \& {Chen}, W. 1997, \apjl, 482, L155, \dodoi{10.1086/310705}

\bibitem[{{Zhang} {et~al.}(2020){Zhang}, {Li}, {Lu}, {Song}, {Xu}, {Liu}, {Chen}, {Cao}, {Bu}, {Chang}, {Chen}, {Chen}, {Chen}, {Chen}, {Chen}, {Cui}, {Cui}, {Deng}, {Dong}, {Du}, {Fu}, {Gao}, {Gao}, {Gao}, {Ge}, {Gu}, {Guan}, {Gungor}, {Guo}, {Han}, {Hu}, {Huang}, {Huo}, {Jia}, {Jiang}, {Jiang}, {Jin}, {Jin}, {Li}, {Li}, {Li}, {Li}, {Li}, {Li}, {Li}, {Li}, {Li}, {Li}, {Li}, {Liang}, {Liao}, {Liu}, {Liu}, {Liu}, {Liu}, {Liu}, {Liu}, {Lu}, {Lu}, {Luo}, {Ma}, {Meng}, {Nang}, {Nie}, {Ou}, {Qu}, {Sai}, {Shang}, {Shen}, {Sun}, {Tan}, {Tao}, {Tuo}, {Wang}, {Wang}, {Wang}, {Wang}, {Wang}, {Wang}, {Wang}, {Wen}, {Wu}, {Wu}, {Wu}, {Xiao}, {Xiong}, {Yan}, {Yang}, {Yang}, {Yang}, {Yi}, {Yuan}, {Zhang}, {Zhang}, {Zhang}, {Zhang}, {Zhang}, {Zhang}, {Zhang}, {Zhang}, {Zhang}, {Zhang}, {Zhang}, {Zhang}, {Zhang}, {Zhang}, {Zhang}, {Zhang}, {Zhang}, {Zhang}, {Zhang}, {Zhang}, {Zhao}, {Zhao}, {Zheng}, {Zhou}, {Zhu}, {Zhu}, {Zhuang}, \& {Insight-HXMT Team}}]{2020Zhang}
{Zhang}, S.-N., {Li}, T., {Lu}, F., {et~al.} 2020, Science China Physics, Mechanics, and Astronomy, 63, 249502, \dodoi{10.1007/s11433-019-1432-6}

\end{thebibliography}
\bibliographystyle{aasjournal}



\end{document}